\documentstyle[prd,eqsecnum,aps,epsf]{revtex}

\newcommand{\beq}{\begin{equation}}
\newcommand{\beqn}{\begin{eqnarray}}
\newcommand{\eeq}{\end{equation}}
\newcommand{\eeqn}{\end{eqnarray}}
\newcommand{\beqa}{\begin{eqnarray}}
\newcommand{\eeqa}{\end{eqnarray}}
\newcommand{\mpl}{m_{\rm pl}}

\newcommand{\lmk}{\left(}
\newcommand{\rmk}{\right)}
\newcommand{\lkk}{\left[}
\newcommand{\rkk}{\right]}
\newcommand{\lnk}{\left\{}

\newcommand{\zk}{z_k}

\newcommand{\ab}{q}

\newcommand{\singlefig}[2]{
\begin{center}
\begin{minipage}{#1}
\epsfxsize=#1
\epsffile{#2}
\end{minipage}
\end{center}}
\newenvironment{figcaption}[2]{
 \vspace{0.3cm}
 \refstepcounter{figure}
 \label{#1}
 \begin{center}
 \begin{minipage}{#2}
 \begingroup \small FIG. \thefigure: }{
 \endgroup
 \end{minipage}
 \end{center}}
\def\beq{\begin{equation}}
\def\eeq{\end{equation}}
\newcommand{\gsim}{\mbox{\raisebox{-1.ex}{$\stackrel
     {\textstyle>}{\textstyle\sim}$}}}
\newcommand{\lsim}{\mbox{\raisebox{-1.ex}{$\stackrel
     {\textstyle<}{\textstyle \sim}$}}}
\newcommand{\square}{\kern1pt\vbox{\hrule height
1.2pt\hbox{\vrule width 1.2pt\hskip 3pt
   \vbox{\vskip 6pt}\hskip 3pt\vrule width 0.6pt}\hrule
height 0.6pt}\kern1pt}

\begin{document}

\title{Cosmological perturbations from multi-field inflation \\
in generalized Einstein theories}
\author{Alexei A. Starobinsky$^{1,2}$, Shinji Tsujikawa$^{2,3}$
and Jun'ichi Yokoyama$^{4}$}

\address{$^1$ Landau Institute for Theoretical Physics,
Russian Academy of Sciences, 117334 Moscow, Russia\\[.3em]}
\address{$^2$ Research Center for the Early Universe,
University of Tokyo, Hongo, Bunkyo-ku, Tokyo 113-0033, Japan}
\address{$^3$ Department of Physics,
Waseda University, 3-4-1 Ohkubo, Shinjuku-ku, Tokyo 169-8555,
Japan\\[.3em]}
\address{$^4$ Department of Earth and Space Science, Graduate School
of Science, Osaka University, Toyonaka 560-0043,
Japan\\[.3em]}
 \date{\today}
 \maketitle
\begin{abstract}
We study cosmological perturbations generated from quantum
fluctuations in multi-field inflationary scenarios in generalized
Einstein theories, taking both adiabatic and isocurvature modes into
account. In the slow-roll approximation, explicit closed-form
long-wave solutions for field and metric perturbations are obtained
by the analysis in the Einstein frame. Since the evolution of
fluctuations depends on specific gravity theories, we make detailed
investigations based on analytic and numerical approaches in four
generalized Einstein theories: the Jordan-Brans-Dicke (JBD) theory,
the Einstein gravity with a non-minimally coupled scalar field, the
higher-dimensional Kaluza-Klein theory, and the $R+R^2$ theory with
a non-minimally coupled scalar field. We find that solutions obtained
in the slow-roll approximation show good agreement with full numerical
results except around the end of inflation. Due to the presence of
isocurvature perturbations, the gravitational potential $\Phi$ and
the curvature perturbations ${\cal R}$ and $\zeta$ do not remain
constant on super-horizon scales. In particular, we find that negative
non-minimal coupling can lead to strong enhancement of ${\cal R}$ in
both the Einstein and higher derivative gravity, in which case it is
difficult to unambiguously decompose scalar perturbations into
adiabatic and isocurvature modes during the whole stage of inflation.
\end{abstract}
\vskip 1pc
\vskip 2pc

\section{Introduction}

The beauty of the inflationary paradigm is that it both 1) explains
why the present-day Universe is approximately homogeneous, isotropic
and spatially flat, so that it may be described by the
Friedmann-Robertson-Walker (FRW) model in the zero approximation,
\cite{inf} and 2) makes detailed quantitative predictions about small
deviations from homogeneity and isotropy including density
perturbations which produce gravitationally bound objects (such as
galaxies, quasars, etc.) and the large-scale structure of the
Universe. It is the latter predictions that make it possible to test
and falsify this paradigm (for each its concrete realization) like
any other scientific hypothesis. Fortunately, all existing and
constantly accumulating data, instead of falsifying, continue to
confirm these predictions (within observational errors). Historically,
among models of inflation making use of a scalar field (called an
{\it inflaton}), the original model or the first-order phase
transition model \cite{oriinf} failed due to the graceful exit
problem, which was taken over by the new \cite{newinf} and the chaotic
\cite{chaoinf} inflation scenarios where the inflaton is slowly
rolling during the whole de Sitter (inflationary) stage. The latter
property was shared by the alternative scenario with higher-derivative
quantum gravity corrections \cite{R2} (where the role of an inflaton
is played by the Ricci scalar $R$) just from the beginning. Note that
a simplified version of this scenario - the $R+R^2$ model - was even
shown to be mathematically equivalent to some specific version of the
chaotic scenario \cite{W84} (see also a review in \cite{Got92}).

Turning to inhomogeneous perturbations on the FRW background, the
inflationary paradigm generically predicts two kinds of them: scalar
perturbations and tensor ones (gravitational waves) which are
generated from quantum-gravitational fluctuations of the inflaton
field and the gravitational field respectively during an inflationary
quasi - de Sitter) stage in the early Universe. The spectrum of
tensor perturbations generated during inflation was first derived in
\cite{St79}, while the correct expression for the spectrum of scalar
perturbations {\em after} the end of inflation was obtained in
\cite{pert}. For completeness, one should also cite two papers
\cite{L80} and \cite{M81} where two important intermediate steps on
the way to the right final answer for scalar perturbations were made,
in particular, in the latter paper the spectrum of scalar
perturbations {\em during} inflation was calculated for the
Starobinsky inflationary model \cite{R2}. In order to obtain a small
enough amplitude of density perturbations in all the above mentioned
slow-roll inflationary models, the inflaton should be extremely
weakly coupled to other fields.  It is therefore not easy to find
sound motivations to have such a scalar field in particle physics
(see, however, \cite{MSYY,LR99}).

Reflecting such a situation, the extended inflation \cite{extinf}
scenario was proposed to revive a GUT Higgs field as the inflaton by
adopting non-Einstein gravity theories. Although the first version of
the inflation model, which considers a first-order phase transition
in the Jordan-Brans-Dicke (JBD) theory \cite{BD}, resulted in failure
again due to the graceful-exit problem \cite{presc}, it triggered
further study of more generic class of inflation models in
non-Einstein theories \cite{soft}, in particular extended chaotic
inflation \cite{Lin90} where both the inflaton and the Brans-Dicke
scalar fields are in the slow rolling regime during inflation. Note
that the natural source of Brans-Dicke-like theories of gravity is
the low-energy limit of the superstring theory \cite{FT85,C85}
with the Brans-Dicke scalar being the dilaton.

Several analyses have been done on the density
perturbations produced in extended new or chaotic inflation models
\cite{BM,Mc,MM,Der,Garcia}, all of which made use of
the constancy of the gauge-invariant quantity $\zeta$ \cite{BST}, or
its equivalent ${\cal R}$~\cite{L80,Lyth85} on super-horizon scales,
and matched them directly to quantum field fluctuations at the moment
of horizon crossing which would be the correct procedure in a
single component inflationary model.
However, in the presence of two sources of quantum fluctuations
(i.e., the inflaton and the Brans-Dicke scalar field),
$\zeta$ does not remain constant during inflation due to the
appearance of isocurvature perturbations \cite{KS}. In such a case,
mixing between adiabatic and isocurvature perturbations
may occur due to ambiguity in the definition of the latter ones (see
the discussion in the Sec.~III~A below).

Note that it is nothing unusual nor unexpected about non-conservation
of the quantities $\zeta$ and ${\cal R}$ even in the long-wave, or
super-horizon, $k\ll aH$ limit since they are not conserved integrals
of motion even in the one-field, $k\ll aH$ case, if we follow the
meaning of this term used in classical mechanics. Here, $a(t)$ is the
FRW scale factor, $H\equiv \dot a/a$, where a dot denotes the time
derivative, and $k=|\bf k|$ is the conserved covariant momentum of a
perturbation Fourier mode. Namely, the ``conservation'' of $\zeta$ in
the latter case is restricted to {\em a part} of initial conditions
for perturbations for which the decaying mode is not strongly
dominating. In the opposite case, since then $\zeta = {\cal O}
(k^2\Phi)$, the $k^2$ term in equations for perturbations, e.g. in
Eq. (\ref{dotBardeen}) below, may not be neglected even in the
$k\ll aH$ case. Of course, any classical dynamical system with $N$
degrees of freedom has exactly $2N$ conserved combinations of its
coordinates and conjugate momenta irrespective of the fact if it is
integrable or not (or even chaotic), but the functional form of these
combinations is generically not universal and strongly depends on
initial conditions. That is why no special attention is paid to such
constant quantities in mechanics. The quantity $\zeta$ is just the
examples of such a conserved combination for the growing mode. For
the purely decaying mode, another conserved combination may be
introduced (see Sec.~III below). It is the specifics of the
inflationary scenario where the case of strongly dominated decaying
mode is excluded that leads to the impression of the universal
conservation of $\zeta$ in the one-field, $k\ll aH$ case. Of
course when more modes for each ${\bf k}$ appear in a multi-field
case, generic non-conservation of $\zeta$ becomes more transparent.
This general remark explains numerous findings of non-conservation of
this quantity in particular cases. However, this circumstance does not
affect the predictive power of the inflationary paradigm at all since
metric perturbations, in particular the gravitational potential
$\Phi$, may be calculated during and after inflation without any
reference to $\zeta$ conservation or non-conservation. The only
problem is that evolution of isocurvature modes of scalar
perturbations (in contrast to adiabatic ones) is not universal, and
its knowledge requires some additional assumptions about behavior of
matter after the end of inflation (in particular, isocurvature modes
disappear completely if the total thermodynamic equilibrium is
reached at some moment of time).


A very important point in the derivation of spectra of primordial
perturbations generated in the inflationary scenario is played by
exact solutions of perturbations equations in the long-wave, or
super-horizon limit $k\to 0$.  It is these solutions that give us a
possibility to match quantum inflaton perturbations inside the de
Sitter horizon during inflation to scalar perturbations during
matter or radiation dominated eras, bypassing the study of
physical processes in the Universe in between. That is why such
solutions were sought and found for more and more complicated cases
in numerous papers. In particular, in \cite{SY}, two of the
present authors made the first correct analysis of the issue in the
case of slow-roll inflation in the original Brans-Dicke gravity,
extending the method used in \cite{St85,PS} to find spectra of all
modes of adiabatic and isocurvature fluctuations. This was further
used to constrain the parameters of general scalar-tensor theories
(e.g., the Damour-Nordvedt model \cite{DN}) from the spectrum of
the CMB anisotropy in \cite{GW1,CSY}. General analytic formula
for evaluating the spectral index in multi-field inflation was
developed in \cite{SS}, which neglected isocurvature modes.
$k\to 0$ solutions for all modes in the case of a factorizable
potential $U=V_1(\varphi_1)V_2(\varphi_2)$ were found in \cite{GW2}.
Finally, a general $k\to 0$ solution for perturbations in the
two-field case with an arbitrary potential $U(\varphi_1, \varphi_2)$
was found in \cite{multiMS} in the form of a functional over an
inflationary background solution, from which explicit solutions for
perturbations can be obtained in any case when the background
solution may be explicitly integrated in the slow-roll approximation.

However, there exists an old and completely different way to
obtain super-horizon ($k\to 0$) solutions for perturbations, even
without writing corresponding equations for them:
the Lagrange method of variation of the background FRW solution
$a(t),\phi_n(t), n=1, ... N$ with respect to all constants
entering in it (here $n$ numerates different scalar fields).
In the case of matter in the form of classical fields, this
method always directly produces all $2N$ physical modes of the
$k\to 0$ solution for some quantities (in particular, for some
gauge-invariant quantities, too), though to obtain the full
$k\to 0$ solution for {\em all} gauge-invariant quantities one has
additionally either to use the $0-i$ Einstein equations, or to
integrate the $i-j~(i\not= j),i,j=1,2,3$ 
Einstein equations.\footnote{For matter in the form of $N$ 
hydrodynamic fluids,
only $N$ non-decreasing modes, including the most important 
growing adiabatic mode, may be obtained in this way.} 
The latter necessity
is related to the fact that all gauge-invariant quantities
constructed from a space-time metric and its first derivatives
(including a Newtonian potential $\Phi$) are necessarily non-local,
in accordance with the Einstein equivalence principle. As a result,
their expressions in terms of quantities which may be defined and
measured locally contain $k^2$ in the denominator. So, the next order
of the series in powers of $k^2$ (beyond that which follows from the
Lagrange method) has to be computed for their determination.

Since the background FRW space-time is in the synchronous form,
the Lagrangian method yields $k\to 0$ perturbations in the
synchronous gauge, too. Of course, this method produces solutions
in the {\em explicit} from only if  the background solution is
known explicitly, too. Thus, its power in the two-field case is
the same as that of the Mukhanov-Steinhardt functional. On the
other hand, it is applicable for any number of scalar fields and
during any stage of the Universe evolution, not necessarily at
slow-roll inflation.

In the case of adiabatic modes of perturbations (both growing and
decaying ones), this method was not only known for many years, but
has been already used by Lifshitz and Khalatnikov in a more
advanced form -- to produce non-homogeneous solutions near
singularity without assuming inhomogeneous perturbations to be
small. This is achieved by taking integration constants in the
background FRW solution as arbitrary functions of spatial
coordinates.  That was how their quasi-isotropic solution~\cite{LH60},
or the so called 7-functional solution~\cite{LH63} were constructed.
In the case of isocurvature modes of perturbations, this method
was recently considered in details in \cite{KH98,ST98,NT98}.

We will use this method in the case of adiabatic modes of
perturbations, since it is especially simple in this case. Also, it
provides a simple reason for the universal constancy of the
properly defined growing adiabatic mode at super-horizon scales
(different from that recently proposed in \cite{WMLL}) which does
not depend on any local physical process, in particular, on
presence or absence of preheating. On the other hand, since the
Lagrange method and the Mukhanov-Steinhardt functional are equally
powerful in the two-field case in the slow-roll approximation (so
that we may use any of them), we will use the latter formula to
derive explicit expressions for non-decreasing isocurvature modes of
perturbations.

At present there are many generalized Einstein theories which can
provide inflationary solutions, e.g, generalized scalar-tensor theory,
Einstein gravity with a non-minimally coupled scalar field,
higher-dimensional Kaluza-Klein theories, $f(R)$ gravity theories.
Making use of the conformal equivalence between these theories,
cosmological perturbations can be analyzed in a unified manner (see,
e.g., \cite{Hwang}). In the present paper we analyze density
perturbations generated in slow-roll inflationary models for a
general class of generalized Einstein theories in the presence of two
scalar fields. We make use of the conformal transformation~
\cite{Maeda89} which transforms the original, or the Jordan frame to
the Einstein frame in which equations are somewhat simpler. In
particular, the $i-j~(i\not= j)$ Einstein equations directly lead
to the universal relation (\ref{equal}) for all models considered
in our paper. If this class of gravity theories is considered as a
low-energy limit of the superstring theory, then the Jordan frame
is also called the string frame.

Though it is always possible to write $k\to 0$ solutions for
perturbations in a functional form, actual evolution of perturbations
depends on specific gravity theories. In the JBD theory, where the
Brans-Dicke parameter is constrained to be $\omega > 3500$ from
observations \cite{omega}, the gravitational potential is dominated
by adiabatic perturbations \cite{SY,CSY}, in which case the variation
of ${\cal R}$ is restricted to be small as we will see later. On the
other hand, it was recently found that negative non-minimal coupling
with a second scalar field other than inflaton can lead to
significant growth of $\zeta$ by the analysis in Jordan frame
\cite{TY}. In this case it is not obvious whether the slow-roll
analysis provides correct amplitudes of field and metric
perturbations, since these exhibit strong enhancement during
inflation by negative instability. In this paper we will investigate
the validity of slow-roll approximations by numerical simulations in
the Einstein frame. We will also work on the evolution of cosmological
perturbations in a higher-dimensional theory and the $R+R^2$ theory
with a non-minimally coupled scalar field. The latter corresponds to
the case where explicit and closed forms of solutions in the
super-horizon limit are obtained by slow-roll analysis in spite of
a coupled form of the effective potential.

The rest of the present article is organized as follows. In Sec.~II
we present the Lagrangian in the Einstein frame and introduce several
generalized Einstein theories which can be recasted to the Lagrangian
by conformal transformations.  Then in Sec.~III basic equations and
closed form solutions for super-horizon perturbations are given. In
Sec.~IV we apply the results of Sec.~III to specific gravity theories,
namely, the JBD theory, the Einstein gravity with a non-minimally
coupled scalar field, the higher-dimensional Kaluza-Klein theory and
the $R^2+(1/2)\xi R\chi^2$ theory. In order to confirm analytic
estimates, we also show numerical results by solving full equations
of motion. We present conclusions and discussions in the final
section.


\section{Inflation in generalized Einstein theories}

Consider the following two-field model with scalar fields
$\varphi_1$ and $\varphi_2$:
\begin{eqnarray}
S = \int d^4 x \sqrt{-g} \left[ \frac{1}{2\kappa^2}R- \frac12
(\nabla\varphi_1)^2 -\frac12 e^{-2F(\varphi_1)}(\nabla\varphi_2)^2 -
U(\varphi_1,\varphi_2) \right],
\label{lagrangian}
\end{eqnarray}
where $\kappa^2/8\pi=G$ is the Newton's gravitational constant,
$F(\varphi_1)$ is a function of $\varphi_1$, and
$U(\varphi_1,\varphi_2)$ is a potential of scalar fields.
Many of the generalized Einstein theories
are reduced to the Lagrangian (\ref{lagrangian}) via conformal
transformations \cite{Maeda89}.  We have the following theories, which
may provide inflationary solutions.
\begin{enumerate}
\item
Theories with a scalar field $\psi$ coupled to gravity whose action
is written by
\begin{eqnarray}
S = \int d^4 x \sqrt{-\hat{g}} \left[ f(\psi)R(\hat{g})- h(\psi)
(\hat{\nabla} \psi)^2 -\frac12 (\hat{\nabla}\phi)^2-V(\phi) \right],
\label{lag1}
\end{eqnarray}
where $V(\phi)$ is a potential of inflaton, $\phi$.
In this work we consider the following theories.
\begin{enumerate}
\item Jordan-Brans-Dicke (JBD) theory with a Brans-Dicke field,
$\psi$ \cite{BD}.
In this case $f$ and $h$ are
\begin{eqnarray}
f=\frac{\psi}{16\pi},~~~~ h=\frac{\omega}{16\pi \psi},
\label{brans}
\end{eqnarray}
where $\omega$ is the Brans-Dicke parameter which is restricted as
$\omega>3500$ from observations \cite{omega}.
Making a conformal transformation,
\begin{eqnarray}
g_{\mu\nu}=\Omega^2\hat{g}_{\mu\nu},
\label{conformal}
\end{eqnarray}
where
\begin{eqnarray}
\Omega^2=\frac{\kappa^2}{8\pi}\psi \equiv \exp\left(\frac{\kappa
\chi}{\sqrt{\omega+3/2}}\right),
\label{conformalfac}
\end{eqnarray}
we obtain the action in the Einstein frame (\ref{lagrangian}) with
replacement,
\begin{eqnarray}
\varphi_1 \to \chi, ~~~\varphi_2 \to \phi,
\label{substitute}
\end{eqnarray}
and
\begin{eqnarray}
F=(\beta/4)\kappa \chi,~~~~U(\chi,\phi)=e^{-\beta\kappa\chi}V(\phi),
\label{brans2}
\end{eqnarray}
with $\beta=\sqrt{8/(2\omega+3)}$.

\item Non-minimally coupled massless scalar field, $\psi$, with an
interaction, $(1/2)\xi R\psi^2$ \cite{TY,SH}.
In this case $f$ and $h$ read
\begin{eqnarray}
f=\frac{1-\xi \kappa^2 \psi^2}{2\kappa^2}, ~~~ h=\frac12.
\label{nonmin1}
\end{eqnarray}
Applying the conformal transformation (\ref{conformal}) with
$\Omega^2=1-\xi\kappa^2\psi^2$ and
defining a new field $\chi$ in order for the kinetic term
to be canonical as
\begin{eqnarray}
\chi=\int \sqrt{\frac{1-(1-6\xi)\xi\kappa^2\psi^2}{(1-\xi\kappa^2
\psi^2)^2}}d\psi,
\label{nonmin2}
\end{eqnarray}
we obtain the action (\ref{lagrangian}) with replacement
(\ref{substitute}) and
\begin{eqnarray}
F= \frac12 \ln |1-\xi\kappa^2\psi^2|,~~~~
U(\chi,\phi)=e^{-4F(\chi)}V(\phi)=\frac{V(\phi)}
{(1-\xi\kappa^2\psi^2)^2}.
\label{nonmin3}
\end{eqnarray}
\end{enumerate}
The induced gravity theory \cite{induced} is also described by the
action, (\ref{lag1}). In this theory the scalar
field $\psi$ has its own potential of the form,
$V(\psi)=(\lambda/8)(\psi^2-\eta^2)^2$ with $f=(\epsilon/2)\psi^2$
and $h=1/2$.

\item
The higher-dimensional theories where the inflaton, $\bar{\phi}$, is
introduced in $N=D+4$ dimensions
\begin{eqnarray}
S = \int d^N x \sqrt{-\bar{g}} \left[\frac{\bar{R}}{2\bar{\kappa}^2} -
\frac12(\bar{\nabla}\bar{\phi})^2-\bar{V}(\bar{\phi})\right],
\label{higher}
\end{eqnarray}
where $\bar{\kappa}^2$ and $\bar{R}$ are the $N$-dimensional
gravitational constant and a scalar curvature, respectively.
We compactify the $N$-dimensional spacetime into the four-dimensional
spacetime and the $D$-dimensional internal space with length scale,
$b$. Then the metric can be expressed as
\begin{eqnarray}
ds_N^2=\hat{g}_{\mu\nu}dx^{\mu}dx^{\nu}
+b^2 ds^2_D,
\label{highermetric}
\end{eqnarray}
where $\hat{g}_{\mu\nu}$ is a four-dimensional metric. Assuming that
extra dimensions are compactified on a torus which has zero curvature,
\footnote{Note that there exist other methods of compactifications.
One of them is the compactification on the sphere \cite{CW}, in which
case stability of extra dimensions and the evolution of
cosmological perturbations during inflation are studied in
\cite{Amendola,shinji00}.}
one gets the following action after dimensional reduction \cite{BM}:
\begin{eqnarray}
S=\int d^4 x \sqrt{-\hat{g}} \left(\frac{b}{b_0}\right)^D
\frac{1}{2\kappa^2}  \Biggl[
\hat{R} +d(d-1)\frac{\partial_{\mu}b \partial_{\nu}b}{b^2}
\hat{g}^{\mu\nu}-\bar{\kappa}^2 \left\{ \frac12
(\hat{\nabla}\hat{\phi})^2-\hat{V}(\hat{\phi}) \right\} \Biggr],
\label{fourdimen}
\end{eqnarray}
where $b_0$ is the present value of $b$, and $\hat{R}$ is the scalar
curvature with respect to $\hat{g}_{\mu\nu}$.
In order to obtain the Einstein-Hilbert action, we make the conformal
transformation (\ref{conformal}) with a conformal factor,
\begin{eqnarray}
\Omega^2=\exp \left(D \frac{\chi}{\chi_0} \right),
\label{Omegahigh}
\end{eqnarray}
where a new scalar field, $\chi$, is defined by
\begin{eqnarray}
\chi = \chi_0 {\rm ln} \left( \frac{b}{b_0} \right),~~~~{\rm with}~~~~
\chi_0= \left[ \frac{D(D+2)}{2\kappa^2} \right]^{1/2}.
\label{chi}
\end{eqnarray}
Then the four-dimensional action in the Einstein frame
can be described as (\ref{lagrangian}) with replacement
(\ref{substitute}) and
\begin{eqnarray}
F=0,~~~~U(\chi,\phi)=\exp \left(-\beta\kappa \chi \right) V(\phi),
~~~~{\rm with}~~~~\beta=\sqrt{\frac{2D}{D+2}}.
\label{FU2}
\end{eqnarray}
Note that when inflaton is introduced in the four-dimensional action
{\it after} compactification,
we find $F=e^{-(\beta/2)\kappa\chi}$ and
$U(\chi,\phi)=e^{-\beta \kappa\chi} V(\phi)$ with
$\beta=\sqrt{8D/(D+2)}$.
In this case, however, we do not have inflationary solutions since
the effective potential does not satisfy the condition: $\beta<
\sqrt{2}$, which is required for power-law inflation to occur (see
the next section).

\item
The $f(R)$ theories where the Lagrangian includes the higher-order
curvature terms, i.e., $\partial f/\partial R$ depends on the scalar
curvature $R$ \cite{Hwang}:
\begin{eqnarray}
S = \int d^4 x \sqrt{-\hat{g}} \left[f(R)-\frac12 (\hat{\nabla}
\chi)^2\right].
\label{FR}
\end{eqnarray}
In this case the conformal factor
\begin{eqnarray}
\Omega^2=2\kappa^2 \left| \frac{\partial f}{\partial R} \right|,
\label{dynamical}
\end{eqnarray}
describes a dynamical freedom in the Einstein-Hilbert action.
Introducing a new scalar field
\begin{eqnarray}
\phi=\sqrt{\frac{3}{2\kappa^2}} {\rm ln}
\left[ 2\kappa^2 \left| \frac{\partial f}{\partial R} \right|
\right],
\label{newphi}
\end{eqnarray}
the action in the Einstein frame is described as (\ref{lagrangian})
with replacement,
\begin{eqnarray}
\varphi_1 \to \phi, ~~~\varphi_2 \to \chi,
\label{substitute2}
\end{eqnarray}
and
\begin{eqnarray}
F=\frac{\kappa\phi}{\sqrt{6}},~~~~U(\phi,\chi)= ({\rm sign})
\exp\left(-\frac{2\sqrt{6}}{3}\kappa \phi \right) \left[\frac{{\rm
(sign)}}{2\kappa^2} R(\phi,\chi) \exp\left(\frac{\sqrt{6}}{3} \kappa
\phi\right)-f(\phi,\chi) \right],
\label{FU}
\end{eqnarray}
where ${\rm sign}=(\partial f/\partial R)/|\partial f/\partial R|$.
For example, in the $R^2$ theory \cite{R2} with a non-minimally
coupled massless
$\chi$ field, i.e.,
\begin{eqnarray}
f(R)=\frac{1}{2\kappa^2}R+\ab R^2-\frac12\xi R\chi^2,
\label{R2FR}
\end{eqnarray}
the effective two-field potential is described as
\begin{eqnarray}
U(\phi,\chi)=\frac{m_{\rm pl}^4}{(32\pi)^2\ab}
e^{-(2\sqrt{6}/3)\kappa\phi}
\left( e^{(\sqrt{6}/3)\kappa\phi}-1+\xi\kappa^2\chi^2\right)^2,
\label{R2potential}
\end{eqnarray}
where we have chosen a positive sign.  In this case the $\phi$ field
behaves as an inflaton and leads to an inflationary expansion of the
Universe.

\end{enumerate}


\section{Cosmological perturbations in two-field inflation}

\subsection{Basic equations and solution for the adiabatic mode}

Let us first derive the background equations. Variation of the action
(\ref{lagrangian}) yields the following background equations for the
cosmic expansion rate $H =\dot{a}/a$ and homogeneous parts
of scalar fields:
\begin{eqnarray}
H^2=\frac{\kappa^2}{3} \left(\frac12 \dot{\varphi}_1^2+\frac12 e^{-2F}
\dot{\varphi}_2^2+U \right),
\label{hubble1}
\end{eqnarray}
\begin{eqnarray}
\dot{H}=-\frac{\kappa^2}{2} \left( \dot{\varphi}_1^2
+e^{-2F}\dot{\varphi}_2^2 \right),
\label{hubble2}
\end{eqnarray}
\begin{eqnarray}
\ddot{\varphi}_1+3H\dot{\varphi}_1+U_{,\varphi_1}
+F_{,\varphi_1}  e^{-2F} \dot{\varphi}_2^2=0,
\label{varphi1}
\end{eqnarray}
\begin{eqnarray}
\ddot{\varphi}_2+3H\dot{\varphi}_2+
e^{2F}U_{,\varphi_2}-2F_{,\varphi_1} 
\dot{\varphi}_1\dot{\varphi}_2=0,
\label{varphi2}
\end{eqnarray}
where a prime denotes a derivative with respect to $\varphi_1$.
A generic solution of this system contains 5 arbitrary
integration constants (two constants appear from the solution of
Eq.~(\ref{varphi1}), two -- from Eq.~(\ref{varphi2}), and one
from Eq.~(\ref{hubble1}), while Eq.~(\ref{hubble2}) is a consequence
of the other equations). However, one of these constants corresponds
to a trivial shift of the cosmic time $t$. The Lagrange variation
with respect to these constant yields a gauge mode. So, variation
with respect to only 4 constants may be used to produce physical
solutions for perturbations in the long-wave limit.

Moving to perturbations now, we restrict ourselves to the
spatially flat FRW background, first, for simplicity and, second,
because recent data on angular fluctuations of the cosmic
microwave background (CMB) convincingly confirm the absence of any
significant spatial curvature of the Universe within a few percent
accuracy (see, e.g., \cite{Boom01}). Then a perturbed space-time
metric has the following form for scalar perturbation in an arbitrary
gauge:
\begin{eqnarray}
ds^2=-(1+2A)dt^2 + 2a(t)B_{,i} dx^idt
+a^2(t)[(1+2D)\delta_{ij}+2E_{,i,j}] dx^i dx^j, ~ i,j=1,2,3,
\label{metric}
\end{eqnarray}
where a comma means usual flat space coordinate derivative and
$\Delta$ is the flat $3D$ Laplacian. In the synchronous gauge,
$A=B=0,~D=(\lambda + \mu)/6,~\Delta E = -\lambda/2$ in the Lifshitz
notations. In the longitudinal gauge, $B=E=0,~A=\Phi,~D=-\Psi$, and
$\Phi,\Psi$ are gauge-invariant potentials \cite{Bardeen,MFB,LL93}.
Further, we assume the $\exp(i{\bf kx})$ dependence for each
Fourier mode ${\bf k}$ and omit the subscript $k\equiv |{\bf k}|$ in
expressions for time-dependent parts of perturbations. Note the
useful relations between $\lambda,\mu$ and $\Phi,\Psi$, and also
between scalar field perturbations in the synchronous gauge
$\delta\varphi_{S,n},~n=1,2$ and in the longitudinal gauge
$\delta\varphi_n$ (the latter quantities are gauge-invariant
actually):
\begin{equation}
\Phi= -{1\over 2k^2}{d\over dt}(a^2\dot\lambda)~,~~~~
\Psi= -{1\over 6}(\lambda+\mu)+{a\dot a\over 2k^2}
\dot\lambda~,~~~~
\delta\varphi_n = \delta\varphi_{S,n} - {\dot\varphi_n\over 2k^2}
\dot\lambda~.
\label{relation}
\end{equation}
Another useful gauge-invariant scalar field perturbation is given
by the Mukhanov-Sasaki variable~\cite{MSvar}:
\begin{equation}
q_n = \delta\varphi_n + {\dot\varphi_n\over H}\Psi =
\delta\varphi_{S,n}-{\dot\varphi_n\over 6H}~(\lambda+\mu)~.
\label{xi}
\end{equation}

For all models considered in our paper, we have the following
relation in the Einstein frame
\begin{eqnarray}
\Phi=\Psi,
\label{equal}
\end{eqnarray}
which follows from the $i-j~(i\not=j)$ Einstein equations taking
into account the fact that anisotropic stresses vanish at the linear
order there. Transforming back to the Jordan frame, this relation
does not hold generically \cite{KS,Hwang}.

Now we can derive one solution for super-horizon perturbations
using the Lagrange method without writing equations for
perturbations. This is possible since one of integration
constants of a background FRW solution, namely, that which appears
by integrating Eq. (\ref{hubble1}) is trivial: it is simply a
multiplier $a_0$ of $a(t)$. Of course, due to invariance of
measurable quantities with respect to equal rescaling of all 3
spatial coordinates $x^i$, this constant does not appear in
variables like $H(t)$ and $\varphi_n(t)$. So, in the $k\to 0$ limit,
\begin{equation}
\mu=6{\delta a_0\over a_0}\equiv 3h={\rm const};~~~\lambda,
\delta\varphi_{S,n}={\cal O}(k^2h);~~~q_n=
-h{\dot\varphi_n\over 2H}.
\label{adia}
\end{equation}
This formula is valid both in the Jordan and the Einstein frames.
By definition, this partial solution will be called the {\em
growing adiabatic mode} (we will discuss the ambiguity of this
definition later). It is clear that this solution exists
for {\em any} form of the gravity Lagrangian and the matter
energy-momentum tensor. In particular, presence of fast oscillations
in a background solution does not affect it, too. The only things
which are needed for its existence are the spatial flatness and
isotropy of the background metric. That is why the derivation
of the spectrum of adiabatic perturbations generated during
inflation in the second of Refs.~\cite{pert}, where this method
was used even in a more general form (valid when $|h|$ is not
necessarily small), does not depend on any physics between
inflation and the present era. Note also that the invariance of
measurable quantities with respect to different rescaling
of spatial coordinates with the total volume fixed leads to
the constancy of the non-decreasing (quasi-isotropic) mode of
gravitational waves in the super-horizon regime. That property
played a crucial role in the derivation of the spectrum of
gravitational waves produced during inflation in~\cite{St79}.

Let us now calculate the gauge-invariant quantities $\Phi$ and
$\delta\varphi_n$ for the solution (\ref{adia}) in the Einstein
frame where the relation (\ref{equal}) holds. In the synchronous
gauge, the latter relation reads
\begin{equation}
\ddot\lambda + 3H\dot\lambda -{k^2\over 3a^2}(\lambda+\mu)=0~.
\end{equation}
Thus, in the $k\to 0$ limit, we get
\begin{equation}
\lambda={h\over a^3}\int_{t_1}^ta\,dt~,~~~\Phi=\Psi=-{h\over 2}
\left(1-{H\over a}\int_{t_1}^ta\,dt\right)~,~~~
\delta\varphi_n=-h{\dot\varphi_n\over 2a}\int_{t_1}^ta\,dt
\label{phiadia}
\end{equation}
($t_1$ depends on ${\bf k}$ generically). A shift in the $t_1$
produces one more superhorizon solution -- the {\em decaying
adiabatic mode}:
\begin{equation}
\lambda=h_1a^{-3},~~~\mu={\cal O}(h_1k^2),~~~\Phi=
\Psi=h_1H/2a,~~~
\delta\varphi_n=-h_1\dot\varphi_n/2a~.
\label{decadia}
\end{equation}
Thus, the growing adiabatic mode (\ref{adia}) is defined up to
an addition of some amount of the decaying adiabatic mode
(\ref{decadia}). In the inflationary scenario, $t_1$ is the moment
of the first Hubble radius crossing during inflation for each
${\bf k}$, so there is no ambiguity at all. Thus, as a whole
the adiabatic solution (mode) is defined unambiguously by Eqs.~
(\ref{adia},~\ref{phiadia}). On the other hand, one may always add
some amount of the adiabatic mode to other, isocurvature solutions
(modes). So, the definition of isocurvature modes is not unique.
In particular, this ambiguity may be used to make $\Phi=\Psi=0$
for an isocurvature mode in the Einstein frame at some chosen
moment of time, e.g., at the end of inflation or at the moment
when the full thermal equilibrium is reached (if the latter occurs
at all). Of course, this choice does not affect any observable
quantities.

A useful gauge-invariant quantity is the comoving curvature
perturbation ${\cal R}$~\cite{L80,Lyth85}:
\begin{equation}
{\cal R}= -{1\over 6}(\lambda+\mu)+{H\over 6\dot H}
(\dot\lambda + \dot\mu)=\Psi-\frac{H}{\dot{H}}(\dot\Psi+H\Phi)~.
\label{calR}
\end{equation}
A similar quantity is the curvature perturbation on uniform
density hypersurfaces introduced in~\cite{BST}:
\begin{equation}
\zeta=D-H{\delta\rho\over\dot\rho}=\left({1\over 6}-{k^2\over
18a^2\dot H}\right)(\lambda+\mu)-{H\over 6\dot H}\dot\mu
=-{\cal R}+{k^2\over 3a^2\dot H}\Psi
\label{Bardeen}
\end{equation}
where $\rho$ is the total energy density of matter and
$\delta\rho$ -- its perturbation. For the solution (\ref{adia}),
we get
\begin{equation}
{\cal R}=-\,\zeta=-h/2={\rm const}\not=0.
\label{constant}
\end{equation}
Thus, we have the theorem: \\
{\it In the superhorizon limit} $k\to 0$ {\it and for $\dot H\not=0$
identically, there always exists one solution for perturbations (the
growing adiabatic mode) for which} ${\cal R}=-\zeta$ {\it is
conserved in the leading order in} $k^2$ {\it (apart from the
vicinity of points where} $\dot H=0$). \\
Let us emphasize that this statement is valid both in the Jordan
and the Einstein frames.

Pathological behavior of ${\cal R}$ and $\zeta$ at the moments
of time when $\dot H=0$ where they diverge (if terms of the next
order in $k^2$ are taken into account) is solely an artifact of
their definition. The potentials $\Phi$ and $\Psi$ remain regular
and small near these points, so no consideration of non-linear
effects is required there. Moreover, it immediately follows from
the formulas (\ref{adia},~\ref{phiadia}) that the same constant
value (\ref{constant}) of ${\cal R}$ and $\zeta$ is restored
after passing through any point where $\dot H=0$. This behavior
of $\zeta$ is clearly seen in our numerical calculations in
Figs. 1 and 3 below.

What about conservation of $\cal R$ and $\zeta$ for the
decaying adiabatic mode (\ref{decadia}) ? Here, there is a subtle
point. In the leading order, ${\cal R}=\zeta=0$ for this mode.
However, one may not say that these quantities are conserved (even
approximately). Real (approximate) conservation would require
$H|\dot {\cal R}/{\cal R}|\ll 1$ that is not valid if the decaying
adiabatic mode is strongly dominating \cite{Leach}. 
Also, the relation ${\cal R}
\approx -\zeta$ is not correct for this mode. On the other hand,
in the Einstein frame one can introduce the gauge invariant quantity
${\cal T}\equiv a\Phi/H$ which does conserve in the super-horizon
limit in the leading order in $k^2$.

Finally, for other solutions (isocurvature modes) ${\cal R}$ and
$\zeta$ are not conserved in the super-horizon limit, too. This can
be easily seen by considering the time derivative of ${\cal R}$
\cite{GW1,GW2}:
\begin{eqnarray}
\dot{{\cal R}}=\frac{H}{\dot{H}}\frac{k^2}{a^2}\Phi
+H\left(\frac{\delta\varphi_1}{\dot{\varphi_1}}-
\frac{\delta\varphi_2}{\dot{\varphi_2}}\right) Z,
\label{dotBardeen}
\end{eqnarray}
where
\begin{eqnarray}
Z \equiv \frac{2e^{-2F}\dot{\varphi}_1\dot{\varphi}_2
(\ddot{\varphi}_1\dot{\varphi}_2-\dot{\varphi}_1\ddot{\varphi}_2
+\dot{F}\varphi_1\varphi_2)+F_{,\varphi_1}
\dot{\varphi}_1e^{-4F}\dot{\varphi}_2^4}{(\dot{\varphi}_1^2+e^{-2F}
\dot{\varphi}_2^2)^2}.
\label{Z}
\end{eqnarray}
The quantity
\begin{eqnarray}
S_{\varphi_1\varphi_2} \equiv
H\left(\frac{\delta\varphi_1}{\dot{\varphi_1}}-
\frac{\delta\varphi_2}{\dot{\varphi_2}}\right),
\label{entropy}
\end{eqnarray}
represents a generalized entropy perturbation between $\varphi_1$ and
$\varphi_2$ fields\cite{GWBM,HN}. In the multi-field case,
$S_{\varphi_1\varphi_2} \ne 0$ and $Z \ne 0$ generically. So, the
presence of isocurvature perturbations leads to the variation of
${\cal R}\approx - \zeta$.

Note also that we use the terms "adiabatic perturbations" and
"isocurvature perturbations", as is commonly done, to denote
different {\em solutions} (modes) of the same physical variables.
This should be contrasted with the approach of 
the recent paper \cite{GWBM} where "adiabatic" 
and "entropy" {\em fields} are
introduced which are certain linear combinations of initial fields
$\varphi_n(t)$, and then adiabatic and entropy perturbations
mean perturbations of these fields. However, for our adiabatic
solution (\ref{adia},~\ref{phiadia}) the perturbed entropy field
$\delta s\propto(\dot\varphi_1\delta\varphi_2-\dot\varphi_2\delta
\varphi_1)=0$ in the super-horizon limit, even if the background
field trajectory in the scalar field space is curved. Thus, the
adiabatic mode is not sourced by isocurvature (entropy) modes, in
contrast to results of \cite{GWBM}.

That is all what may be obtained without solving equations for
perturbations. So, to proceed further, equations for time-dependent
parts of metric and field fluctuations have to be written. They
read:
\begin{eqnarray}
\ddot{\Phi}+4H\dot{\Phi}+\kappa^2U\Phi= \frac{\kappa^2}{2} 
\left[ \dot{\varphi}_1\delta\dot{\varphi}_1-
(U_{,\varphi_1}+F_{,\varphi_1} 
e^{-2F}\dot{\varphi_2}^2) \delta\varphi_1+
e^{-2F}\dot{\varphi}_2\delta\dot{\varphi}_2-
U_{,\varphi_2}\delta\varphi_2 \right],
\label{perturb4}
\end{eqnarray}
\begin{eqnarray}
\left(\frac{k^2}{a^2}-\dot{H}\right)\Phi
=-\frac{\kappa^2}{2}\left[\dot{\varphi}_1\delta\dot{\varphi}_1
+(3H\dot{\varphi}_1+U_{,\varphi_1}-F_{,\varphi_1}e^{-2F}
\dot{\varphi}_2^2) \delta\varphi_1+
e^{-2F}\dot{\varphi}_2\delta\dot{\varphi}_2+
(U_{,\varphi_2}+3H\dot{\varphi}_2e^{-2F}) \delta\varphi_2 \right],
\label{perturb5}
\end{eqnarray}
\begin{eqnarray}
\dot{\Phi}+H\Phi=\frac{\kappa^2}{2}
\left( \dot{\varphi}_1 \delta \varphi_1
+e^{-2F}\dot{\varphi}_2 \delta \varphi_2 \right),
\label{perturb1}
\end{eqnarray}
\begin{eqnarray}
\delta\ddot{\varphi}_1&+& 3H\delta\dot{\varphi}_1
+\left[\frac{k^2}{a^2}+U_{,\varphi_1\varphi_1}-
\left(e^{-2F}\right)_{,\varphi_1\varphi_1}
 \frac{\dot{\varphi}_2^2}{2} \right] \delta\varphi_1
+ 2F_{,\varphi_1} e^{-2F} \dot{\varphi}_2
\delta\dot{\varphi}_2+U_{,\varphi_1\varphi_2}\delta\varphi_2 \nonumber \\
&=& 4\dot{\varphi}_1 \dot{\Phi}-2U_{,\varphi_1}\Phi,
\label{perturb2}
\end{eqnarray}
\begin{eqnarray}
\delta\ddot{\varphi}_2 &+& (3H-2\dot{F})\delta\dot{\varphi}_2
+\left(\frac{k^2}{a^2}+e^{2F}U_{,\varphi_2\varphi_2} \right)
\delta\varphi_2-2\dot{F}\delta\dot{\varphi}_1+e^{2F}
\left( 2F_{,\varphi_1}
U_{\varphi_2}+U_{\varphi_1 \varphi_2} \right) \delta \varphi_1
\nonumber \\
&=& 4\dot{\varphi}_2 \dot{\Phi}-2e^{2F}U_{,\varphi_2}\Phi.
\label{perturb3}
\end{eqnarray}
The relation (\ref{perturb5}) clearly indicates that metric
perturbations are determined when the evolution of scalar fields are
known.

\subsection{Closed form solutions in slow-roll approximations}

The use of the slow-roll approximation allows us to obtain closed form
solutions for isocurvature perturbations in the long-wave limit
\cite{SY,GW1,GW2,multiMS}.
Under this approximation, the background equations are simplified as
\begin{eqnarray}
H^2=\frac{\kappa^2}{3} U,
\label{hubbleslow}
\end{eqnarray}
\begin{eqnarray}
3H\dot{\varphi}_1+U_{,\varphi_1}=0,
\label{varphi1slow}
\end{eqnarray}
\begin{eqnarray}
3H\dot{\varphi}_2+e^{2F}U_{,\varphi_2}=0.
\label{varphi2slow}
\end{eqnarray}
Combining Eqs.~(\ref{hubbleslow})-(\ref{varphi2slow}) with
Eq.~(\ref{hubble2}), we find
\begin{eqnarray}
\dot{\varphi}_1=-\frac{H}{\kappa^2}
\frac{U_{,\varphi_1}}{U},~~~~~
\dot{\varphi}_2=-\frac{He^{2F}}{\kappa^2}
\frac{U_{,\varphi_2}}{U},
\label{varphiS}
\end{eqnarray}
\begin{eqnarray}
-\frac{\dot{H}}{H^2}=\frac{1}{2\kappa^2}
\left[\left(\frac{U_{,\varphi_1}}{U} \right)^2+
e^{2F}\left(\frac{U_{,\varphi_2}}{U} \right)^2 \right].
\label{dothubble}
\end{eqnarray}

In JBD and higher-dimensional theories where the potentials take the form,
$U(\varphi_1, \varphi_2)= e^{-\beta \kappa \varphi_1}V(\varphi_2)$,
it is straightforward to show that the scale factor evolves as power-law.
In this case integrating $\dot{\varphi}_1=\beta H/\kappa$ over $t$, we find
\begin{eqnarray}
\varphi_1(t)=\frac{\beta}{\kappa} {\rm ln} \frac{a(t)}{a_f}+\varphi_{1f}
\equiv -\frac{\beta}{\kappa}z+\varphi_{1f},
\label{backphi1}
\end{eqnarray}
where a subscript $f$ denotes the value of each quantity at the end of
inflation and $z$ is the number of $e$-folds of inflationary expansion
after the time $t$.
Assuming that $V(\varphi_2)$
takes a constant value $V_0$ during inflation, one finds
\begin{eqnarray}
a=a_0 \left[ \left\{ \frac{\kappa^2}{3}e^{-\beta\kappa\varphi_1(0)}
V_0\right\}^{1/2} \frac{\beta^2}{2} (t-t_0) \right]^{2/\beta^2},
\label{powerlaw}
\end{eqnarray}
where $a_0$ and $t_0$ are constants.
Then we have a power-law inflationary solution when $\beta<\sqrt{2}$.
For example, in the JBD case with potential, (\ref{brans2}), inflation
takes place with a large power exponent because $\beta$ is constrained
to be $\beta=\sqrt{8/2\omega+3}~\lsim~0.034$ from observations
\cite{omega}. In higher-dimensional theories with potential
(\ref{FU2}), inflation is realized for arbitrary extra dimensions $D$,
because the condition, $\beta<\sqrt{2}$, always holds.

Let us consider large scale perturbations with $k \ll aH$.  Neglecting
$\dot{\Phi}$ and those terms which include second order time
derivatives in Eqs.~(\ref{perturb1})-(\ref{perturb3}), one finds
\begin{eqnarray}
\Phi=\frac{\kappa^2}{2H}
\left( \dot{\varphi}_1 \delta \varphi_1
+e^{-2F}\dot{\varphi}_2 \delta \varphi_2 \right),
\label{perturb1slow}
\end{eqnarray}
\begin{eqnarray}
3H\delta\dot{\varphi}_1+U_{,\varphi_1\varphi_1}
\delta\varphi_1
+U_{,\varphi_1\varphi_2}\delta\varphi_2
+2U_{,\varphi_1}\Phi=0,
\label{perturb2slow}
\end{eqnarray}
\begin{eqnarray}
3H\delta\dot{\varphi}_2+\left(e^{2F}U_{,\varphi_2}\right)_{,\varphi_1}
\delta\varphi_1+\left(e^{2F}U_{,\varphi_2}\right)_{,\varphi_2}
\delta\varphi_2
+2e^{2F}U_{,\varphi_2}\Phi=0.
\label{perturb3slow}
\end{eqnarray}
Note that this approximation may not be always valid especially when
perturbations exhibit nonadiabatic growth during inflation.  We will check
its validity in the next section.

In the slow-roll approximation, the expression (\ref{phiadia}) for the
growing adiabatic mode is simplified:
\begin{equation}
\Phi=\Psi=h{\dot H\over 2H^2}~,~~\delta\varphi_n=-h{\dot\varphi_n\over 2H}~.
\end{equation}
To find a generic solution, we introduce new variables, $x$ and $y$, with
$\delta\varphi_1=U_{,\varphi_1}x$ and $\delta \varphi_2=
e^{2F}U_{,\varphi_2}y$. Then
Eqs.~(\ref{perturb2slow}) and (\ref{perturb3slow}) yield
\begin{eqnarray}
3H\dot{x}+\frac{U_{,\varphi_1\varphi_2} U_{,\varphi_2}e^{2F}}{U_{,\varphi_1}}
(y-x)+2\Phi=0,
\label{x}
\end{eqnarray}
\begin{eqnarray}
3H\dot{y}+\frac{(e^{2F}U_{,\varphi_2})_{,\varphi_1} U_{,\varphi_1}e^{-2F}}
{U_{,\varphi_2}}(x-y)+2\Phi=0.
\label{y}
\end{eqnarray}
Subtracting Eq.~(\ref{x}) from Eq.~(\ref{y}), we obtain the following
integrated solution
\begin{eqnarray}
y-x=Q_3\exp \left[\int \frac{A}{3H}dt\right],
\label{y_x}
\end{eqnarray}
where $Q_3$ is a constant, and
\begin{eqnarray}
A \equiv \frac{U_{,\varphi_1\varphi_2} U_{,\varphi_2}e^{2F}}{U_{,\varphi_1}}
+\frac{(e^{2F}U_{,\varphi_2})_{,\varphi_1}U_{,\varphi_1}e^{-2F}}
{U_{,\varphi_2}}.
\label{A}
\end{eqnarray}
Taking notice of the relation
\begin{eqnarray}
\Phi=\frac{\kappa^2}{2H}
\left[ \dot{U}x+U_{,\varphi_2}\dot{\varphi}_2(y-x)\right]
=\frac{\kappa^2}{2H}
\left[ \dot{U}y+U_{,\varphi_1}\dot{\varphi}_1(x-y)\right],
\label{Phislow}
\end{eqnarray}
and making use of Eq.~(\ref{y_x}) and background equations
(\ref{hubbleslow})-(\ref{varphi2slow}),
we find
\begin{eqnarray}
x &=& -\frac{Q_3}{U} \int \left[ \frac{H}{\kappa^2}
\frac{U_{,\varphi_1\varphi_2} U_{,\varphi_2}e^{2F}}{U_{,\varphi_1}}
+U_{,\varphi_2}\dot{\varphi}_2 \right] \frac{J}{U}dt, \\
y &=& +\frac{Q_3}{U} \int \left[ \frac{H}{\kappa^2}
\frac{(e^{2F}U_{,\varphi_2})_{,\varphi_1} U_{,\varphi_1}}{U_{,\varphi_2}}
+U_{,\varphi_1}\dot{\varphi}_1 \right] \frac{J}{U}dt,
\label{xandy}
\end{eqnarray}
where
\begin{eqnarray}
J \equiv U\exp \left[\int \frac{A}{3H}dt\right].
\label{G}
\end{eqnarray}
Then the final closed-form solutions for long-wave perturbations are
expressed as
\begin{eqnarray}
\delta \varphi_1 = ( {\rm ln} U)_{,\varphi_1}
\left[Q_1+Q_3 \int_{t_*}^{t} \left({\rm ln} ({\rm ln} U)_{,\varphi_1}
\right)_ {,\varphi_2} J d\varphi_2  \right],
\label{finalslow1}
\end{eqnarray}
\begin{eqnarray}
\delta \varphi_2 = e^{2F} ( {\rm ln} U)_{,\varphi_2}
\left[Q_2-Q_3 \int_{t_*}^{t} \left({\rm ln} \left( e^{2F}({\rm ln} U)_
{,\varphi_2} \right) \right)_{,\varphi_1}
 J d\varphi_1  \right],
 \label{finalslow2}
 \end{eqnarray}
\begin{eqnarray}
 \Phi = -\frac12 \left[({\rm ln} U)_{,\varphi_1} \delta\varphi_1+
 ({\rm ln} U)_{,\varphi_2} \delta\varphi_2 \right],
\label{finalslow3}
\end{eqnarray}
with
\begin{eqnarray}
J=\exp \left\{ -\int_{t_*}^{t} \left[  \left({\rm ln}
\left( e^{2F}({\rm ln} U)_{,\varphi_2} \right) \right)_{,\varphi_1} d\varphi_1+
\left({\rm ln} ({\rm ln} U)_{,\varphi_1}
\right)_ {,\varphi_2} d\varphi_2 \right] \right\}.
\label{G2}
\end{eqnarray}
Here integration constants, $Q_1$, $Q_2$, and $Q_3$ satisfy the relation
$Q_2=Q_1+Q_3$, which comes from  Eq.~(\ref{y_x}).
The $Q_3$ terms appear due to the presence of
isocurvature perturbations.
These constants are evaluated by the amplitudes
of quantum fluctuations of scalar fields at horizon crossing, $t_*$.
The fluctuations are generated by small scale perturbations ($k>aH$), so
that they can be considered as free massless scalar fields which are
described by independent random variables \cite{SY,PS}.
Then the field
perturbations when they crossed the Hubble radius ($k \simeq aH$)
are written in the form:
\begin{eqnarray}
\delta \varphi_1({\bf k})|_{t=t_*}=\frac{H(t_*)}{\sqrt{2k^3}}
e_{\varphi_1}({\bf k}),~~~~
\delta \varphi_2({\bf k})|_{t=t_*}=\frac{H(t_*)}{\sqrt{2k^3}}
e^{F(t_*)} e_{\varphi_2}({\bf k}).
\label{quantum}
\end{eqnarray}
Here $e_{\varphi_1}$ and $e_{\varphi_2}$ are classical stochastic Gaussian
quantities, described by
\begin{eqnarray}
\langle e_{\varphi_1} ({\bf k}) \rangle =
\langle e_{\varphi_2} ({\bf k}) \rangle=0,~~~~
\langle e_{\varphi_1} ({\bf k}) e^*_{\varphi_2}({\bf k'})
\rangle=\delta_{ij}\delta^{(3)}({\bf k}-{\bf k'}),
\label{epsilon}
\end{eqnarray}
where $i, j=\varphi_1, \varphi_2$.
{}From Eqs.~(\ref{varphiS}), (\ref{finalslow1}), and (\ref{finalslow2}),
we find
\begin{eqnarray}
\frac{\delta \varphi_1}{\dot{\varphi}_1}=-\frac{\kappa^2}{H}
\left[Q_1+Q_3 \int_{t_*}^{t} \left({\rm ln} ({\rm ln} U)_{,\varphi_1}
\right)_ {,\varphi_2} J d\varphi_2  \right],
\label{ratioslow1}
\end{eqnarray}
\begin{eqnarray}
\frac{\delta \varphi_2}{\dot{\varphi}_2}=-\frac{\kappa^2}{H}
\left[Q_1+Q_3-Q_3 \int_{t_*}^{t} \left({\rm ln} \left( e^{2F}({\rm ln} U)_
{,\varphi_2} \right) \right)_{,\varphi_1}
 J d\varphi_1  \right].
 \label{ratioslow2}
 \end{eqnarray}
Making use of Eqs.~(\ref{quantum}), (\ref{ratioslow1}), and
(\ref{ratioslow2}), the integration constants are expressed as
\begin{eqnarray}
Q_1=-\frac{H^2(t_*)}{\kappa^2\sqrt{2k^3}}
\left( \frac{e_{\varphi_1}({\bf k})}
{\dot{\varphi}_1}\right)_{t_*},~~~~
Q_3=\frac{H^2(t_*)}{\kappa^2\sqrt{2k^3}} \left( \frac{e_{\varphi_1}({\bf k})}
{\dot{\varphi}_1}- e^{F}\frac{e_{\varphi_2}({\bf k})}{\dot{\varphi}_2}
\right)_{t_*}.
\label{Qvalue}
\end{eqnarray}
In the next section, we will investigate the evolution of large-scale
perturbations in specific gravity theories.

\section{Applications to specific gravity theories}

Among the generalized Einstein theories which we presented in Sec.~II,
most of them take the following separated form of potentials except for the
$f(R)$ theories:
\begin{eqnarray}
U(\varphi_1, \varphi_2)=V_1(\varphi_1)V_2(\varphi_2).
\label{poten}
\end{eqnarray}
In this case, we have an integration constant by making use of
Eq.~(\ref{varphiS}):
\begin{eqnarray}
C=-\kappa^2\int \frac{V_1}{V_{1,\varphi_1}}e^{2F}d\varphi_1+
\kappa^2\int \frac{V_2}{V_{2,\varphi_2}} d\varphi_2.
\label{C}
\end{eqnarray}
This characterizes the trajectory in field space in two-field inflation.

Since $J=e^{-2(F-F_*)}$ in the separated potential, (\ref{poten}),
Eqs.~(\ref{finalslow1}) and (\ref{finalslow2}) are easily integrated to
give
\begin{eqnarray}
\delta\varphi_1=\frac{V_1'}{V_1} Q_1,~~~~~~
\delta\varphi_2=\frac{V_2'}{V_2}
\left[ Q_1 e^{2F}+Q_3 e^{2F_*} \right],
\label{SEPfield}
\end{eqnarray}
together with the gravitational potential,
\begin{eqnarray}
\Phi=-\frac12 \left(\frac{V_1'}{V_1}\right)^2 
Q_1-\frac12 \left(\frac{V_2'}{V_2}\right)^2
\left( Q_1 e^{2F}+Q_3 e^{2F_*} \right).
\label{SEPmet}
\end{eqnarray}
Introducing new integration constants,
$C_1 \equiv -\kappa^2(Q_1+Q_3e^{2F_*})$ and
$C_3 \equiv -\kappa^2Q_3e^{2F_*}$, and making use of
Eq.~(\ref{dothubble}), one finds
\begin{eqnarray}
\delta \varphi_1=-(C_1-C_3)\frac{V_1'}{\kappa^2V_1},~~~~~~~
\delta \varphi_2=-\left[C_1e^{2F}-C_3(e^{2F}-1) \right]
\frac{V_2'}{\kappa^2V_2},
\label{fieldper}
\end{eqnarray}
\begin{eqnarray}
\Phi =C_1 \left( \epsilon_{\varphi_1}+e^{2F}\epsilon_{\varphi_2} 
\right)-C_3 \left[\epsilon_{\varphi_1}+ 
(e^{2F}-1) \epsilon_{\varphi_2} \right]
=-C_1\frac{\dot{H}}{H^2}-C_3
\left[\epsilon_{\varphi_1}+(e^{2F}-1) \epsilon_{\varphi_2} \right],
\label{SEPmet2}
\end{eqnarray}
where $\epsilon_{\varphi_1}$ and $\epsilon_{\varphi_2}$ are given by
\begin{eqnarray}
\epsilon_{\varphi_1} \equiv \frac{1}{2\kappa^2}
\left(\frac{V_1'}{V_1}\right)^2,~~~~
\epsilon_{\varphi_2} \equiv \frac{1}{2\kappa^2}
\left(\frac{V_2'}{V_2}\right)^2.
\label{slowpara}
\end{eqnarray}
{}From Eq.~(\ref{Qvalue}) $C_1$ and $C_3$ are expressed as
\begin{eqnarray}
C_1=\frac{H^2(t_*)}{\sqrt{2k^3}}
\left[ (1-e^{2F})\frac{e_{\varphi_1}({\bf k})}{\dot{\varphi}_1}
+e^{3F} \frac{e_{\varphi_2}({\bf k})}{\dot{\varphi}_2} \right]_ {t_*},
~~~~~C_3=\frac{H^2(t_*)}{\sqrt{2k^3}}
\left[e^{3F}\frac{e_{\varphi_2}({\bf k})}{\dot{\varphi}_2}-
e^{2F}\frac{e_{\varphi_1}({\bf k})}{\dot{\varphi}_1} \right]_{t_*}.
\label{NewC}
\end{eqnarray}
The gravitational potential can also be decomposed in a different way.
For example, let us introduce the integration constants $\tilde{C}_1$ and
$\tilde{C}_3$, defined by
\begin{eqnarray}
\tilde{C}_1 &\equiv& -\kappa^2 Q_1-\frac{\kappa^2 e^{2(F_*-F_f)}}
{1+\alpha_f}Q_3=\frac{H^2(t_*)}{\sqrt{2k^3}}
\left[\left(1-\frac{e^{2(F-F_f)}}{1+\alpha_f} \right)
\frac{e_{\varphi_1}({\bf k})}{\dot{\varphi}_1}
+ \frac{e^{3F-2F_f}}{1+\alpha_f}
\frac{e_{\varphi_2}({\bf k})} {\dot{\varphi}_2} \right]_ {t_*},
\nonumber \\
\tilde{C}_3 &\equiv& -\kappa^2 e^{2(F_*-F_f)}Q_3=
\frac{H^2(t_*)}{\sqrt{2k^3}} \left[ e^{3F-2F_f}
\frac{e_{\varphi_2}({\bf k})}{\dot{\varphi}_2}-
e^{2(F-F_f)} \frac{e_{\varphi_1}({\bf k})}{\dot{\varphi}_1}
\right]_ {t_*},
\label{tildeC}
\end{eqnarray}
where the subscript $f$ denotes the value at the end of inflation, and
\begin{eqnarray}
\alpha \equiv \frac{\epsilon_{\varphi_1}}
{e^{2F}\epsilon_{\varphi_2}}.
\label{alpha}
\end{eqnarray}
Then Eq.~(\ref{SEPmet}) reads
\begin{eqnarray}
\Phi=-\tilde{C}_1\frac{\dot{H}}{H^2}-\tilde{C}_3
\left[ \frac{\epsilon_{\varphi_1}}{1+\alpha_f}+
\left(\frac{e^{2F}}{1+\alpha_f}-e^{2F_f}\right)
\epsilon_{\varphi_2} \right].
\label{SEPmet3}
\end{eqnarray}
This decomposition corresponds to the case where the second
term in the rhs of Eq.~(\ref{SEPmet3}) vanishes at the end
of inflation.

During slow-roll we have that
$|\dot{H}/H^2| \ll 1$ in Eq.~(\ref{dothubble}), which yields from
Eqs.~(\ref{calR}), (\ref{SEPmet2}), and (\ref{SEPmet3}) as
\begin{eqnarray}
{\cal R} \simeq -\frac{H^2}{\dot{H}}\Phi
&=&
C_1-C_3 \frac{\epsilon_{\varphi_1}+(e^{2F}-1)
\epsilon_{\varphi_2}}
{\epsilon_{\varphi_1}+e^{2F}\epsilon_{\varphi_2}}
\label{SEPzeta_m1}\\
&=& \tilde{C_1}-\tilde{C_3} \frac{\epsilon_{\varphi_1}
+\left\{e^{2F}-(1+\alpha_f) e^{2F_f} \right\}
\epsilon_{\varphi_2}}
{(1+\alpha_f)(\epsilon_{\varphi_1} +
e^{2F}\epsilon_{\varphi_2})}.
\label{SEPzeta}
\end{eqnarray}
Note that both decompositions coincide each other in the limit
$\alpha_f \to 0$ and $F_f \to 0$.  When $\alpha_f$ and  $F_f$
are nonvanishing, the second term in Eq.~(\ref{SEPmet2}) gives
contributions to the gravitational potential, $\Phi$.  When this term is
negligible relative to $-C_1\dot{H}/H^2$
during the whole stage of inflation,
the first and second terms in Eq.~(\ref{SEPmet2})
can be identified as adiabatic and isocurvature modes, respectively.
However, in some generalized Einstein theories which we discuss
in the following
subsections, the final $\Phi$ is dominated by the second term in
Eq.~(\ref{SEPmet2}).  In those cases we can no longer regard the second
term as the isocurvature mode at the end of inflation.

The relation (\ref{SEPzeta_m1}) or (\ref{SEPzeta}) indicates
that ${\cal R}$ is not generally conserved during inflation
when both fields are evolving due to
the presence of isocurvature perturbations.  The generalized entropy
perturbations are written as
\begin{eqnarray}
S_{\varphi_1\varphi_2}=-C_3e^{-2F}=-\tilde{C}_3e^{2(F_f-F)}.
\label{ent2}
\end{eqnarray}
Since $S_{\varphi_1\varphi_2}$ and $Z$ do not vanish generally, these
work as a source term for the change of ${\cal R}$. The evolution of
${\cal R}$ depends on specific gravity theories as we will show below.

%
\subsection{JBD theory}

Let us first apply the results in the previous section to the JBD theory
with Eq.~(\ref{brans2}).  In this case Eqs.~(\ref{SEPmet2}) and
(\ref{SEPzeta_m1}) are
\begin{eqnarray}
\Phi=-C_1\frac{\dot{H}}{H^2}-C_3\left[\frac{\beta^2}{2}+
(e^{(\beta/2)\kappa\chi}-1) \epsilon_{\phi}\right],
\label{JBDmetric}
\end{eqnarray}
\begin{eqnarray}
{\cal R}=C_1-C_3\left[1-\frac{1}{e^{(\beta/2)\kappa\chi}
+\beta^2/(2\epsilon_{\phi})}\right].
\label{JBDzeta}
\end{eqnarray}

In the JBD theory, $\beta$ is required to be $\beta~\lsim~0.034$ from
observational constraints, which yields $\epsilon_{\chi}=\beta^2/2 \ll 1$.
In addition to this, the value of $\chi$ should be practically vanishing,
$\chi_f \simeq 0$, at the end of
inflation in order to reproduce the present value of the gravitational
constant, because $\chi$ generally grows only as logarithm of $t$ after
inflation and is even constant during radiation-dominated stage in this
theory.  These lead to $F_f=(\beta/4)\kappa \chi_f \simeq 0$
and $\alpha_f=e^{-2F_f}\epsilon_{\chi_f}/\epsilon_{\phi_f} \simeq 0$,
which implies that the decomposition of Eq.~(\ref{SEPmet3}) looks almost
the same form as that of (\ref{SEPmet2}).
Since the second term in the rhs
of Eq.~(\ref{JBDmetric}) is negligible relative to the first term after
inflation, the term proportional to $C_1$ represents the growing adiabatic
mode \cite{MFB}, while the one proportional to $C_3$ corresponds to the
isocurvature mode \cite{SY,CSY}.  During inflation, ${\cal R}$ evolves due to
the change of the term,
\begin{eqnarray}
W \equiv e^{(\beta/2)\kappa\chi}+\beta^2/(2\epsilon_{\phi}).
\label{W}
\end{eqnarray}

As a specific model of inflation,
 let us first consider chaotic inflation driven by a potential,
$V(\phi)=\lambda_{2n}\phi^{2n}/(2n)$.
In this model, inflation occurs at large $\phi$ and it is terminated
when $|\dot{\phi}/\phi|$ becomes as large as $H$ at $\phi=\phi_f=
\sqrt{2n}/\kappa$.
Making use of Eqs.~(\ref{backphi1}) and (\ref{C}) we find
\begin{eqnarray}
\kappa^2\int_{\phi_f}^{\phi}\frac{V(\varphi)}{V'(\varphi)}d\varphi
 =\frac{\kappa^2}{4n}\lmk\phi^2-\frac{2n}{\kappa^2}\rmk
 =  \frac{2}{\beta^2}(1-e^{\beta\kappa\chi/2})
 = \frac{2}{\beta^2}(1-e^{-\beta^2z/2}) \simeq z.
\end{eqnarray}
The error in the last expression is less than $1.7\%$ for  $z \leq 60$ and
$\beta \leq 0.034$.
Then Eq.~(\ref{W}) is rewritten as
\begin{eqnarray}
W=\left(1-\frac{2}{n}\right)e^{(\beta/2)\kappa\chi}
+\frac{2}{n}+\frac{\beta^2}{2n}.
\label{W2}
\end{eqnarray}
Since $W$ is constant for $n=2$, ${\cal R}$ is conserved
in the quartic potential, $V(\phi)=\lambda_4 \phi^4/4$.
In other cases such as the quadratic potential, ${\cal R}$ evolves
during inflation due to the presence of isocurvature perturbations,
although its change is typically small.
At the end of inflation, ${\cal R}$ takes almost constant value,
${\cal R} \simeq C_1$.

\begin{figure}
\begin{center}
\singlefig{12cm}{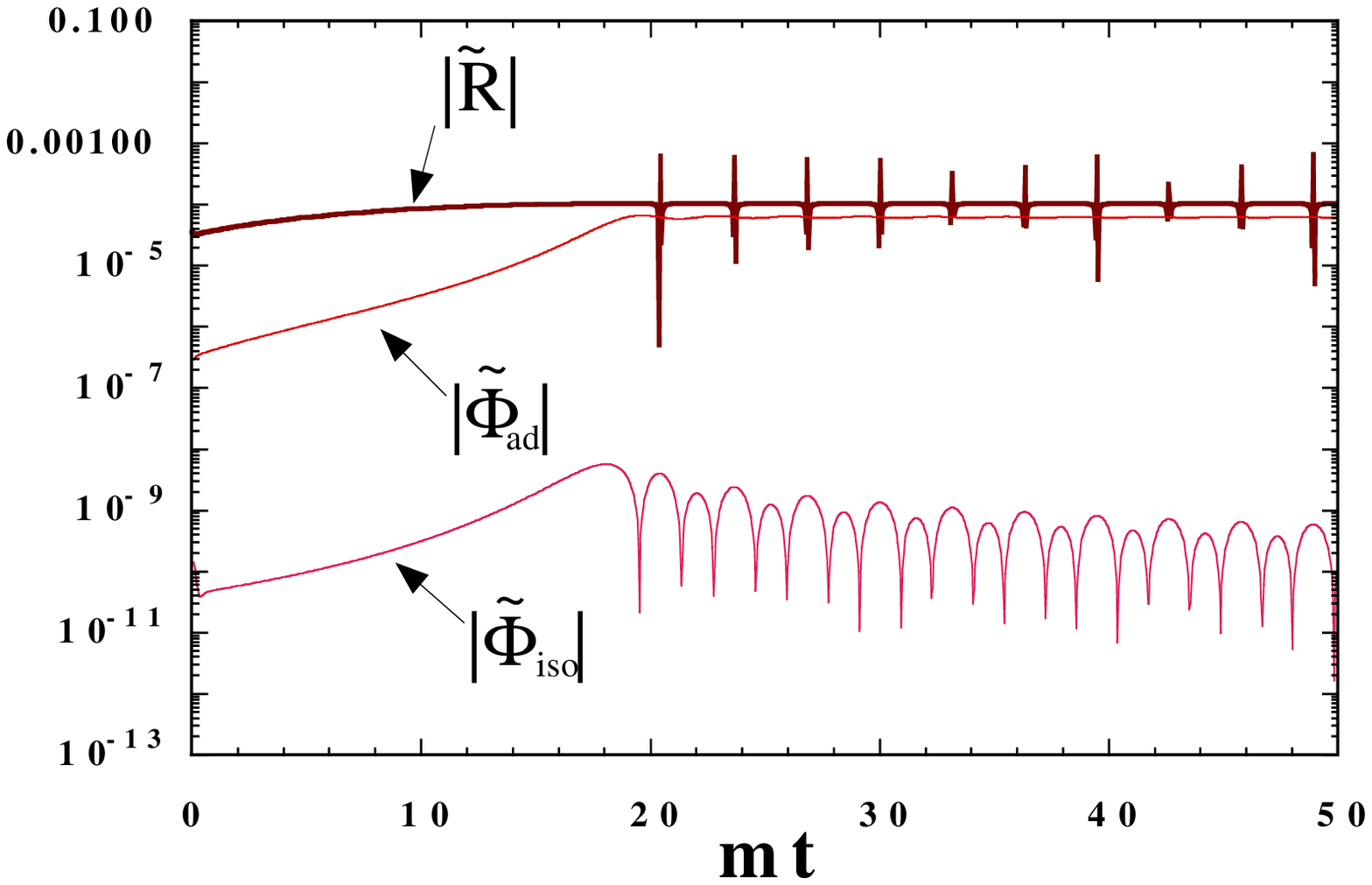}
\begin{figcaption}{Fig1}{12cm}
The evolution of the adiabatic metric perturbation
$\tilde{\Phi}_{\rm ad} \equiv k^{3/2} \Phi_{\rm ad}$,
the isocurvature metric perturbation
$\tilde{\Phi}_{\rm iso} \equiv k^{3/2} \Phi_{\rm iso}$,
and the curvature perturbation $\tilde{\cal R} \equiv k^{3/2} 
{\cal R}$ for the case of $\beta=0.09$
in the JBD theory with quadratic inflaton potential,
$V(\phi)=m^2\phi^2/2$.
We choose the initial values of scalar fields as $\phi=3m_{\rm pl}$
and $\chi=-1.2m_{\rm pl}$, in which case $\chi$ is nearly zero
at the end of inflation.
\end{figcaption}
\end{center}
\end{figure}

One may worry that the above results are obtained
by imposing slow-roll conditions, which are only approximations to the
full equations of motion.
In order to answer such suspicions, we numerically solved full equations
(\ref{perturb4})-(\ref{perturb3}) along with the background equations
(\ref{hubble1})-(\ref{varphi2}).
We adopt the quadratic potential $V(\phi)=m^2\phi^2/2$,
and start integrating from
about  60 e-folds before the end of inflation.
We found that the evolution of field and metric perturbations are well
described by analytic estimations except around the end of inflation.  The
evolution of $\Phi$ for the adiabatic and isocurvature mode is plotted in
Fig.~\ref{Fig1}, which shows that the
contribution of the isocurvature mode is small as estimated by
Eq.~(\ref{JBDmetric}).  The adiabatic growth of the gravitational potential
terminates for $mt~\gsim~20$, after which the system enters a reheating
stage.  During reheating no additional growth of super-Hubble metric
perturbations occurs in this scenario, unless some interactions between
inflaton and other field, $\sigma$, such as $g^2\phi^2\sigma^2/2$ are
not introduced.

The evolution of ${\cal R}$ for $\beta=0.09$
(corresponding to $\omega=500$) is plotted in Fig.~1,
which shows that the change of ${\cal R}$ is small during inflation.
We have also confirmed that the variation 
of ${\cal R}$ is negligible in the case of $\beta~\lsim~0.034$.
Since its growth is sourced by the $e^{(\beta/2)\kappa\chi}$ term in
Eq.~(\ref{W2}), we find in Fig.~1 that ${\cal R}$ approaches 
almost constant value around the end of inflation.  
During reheating ${\cal R}$ is conserved
except for the short period when $\dot{H}$ passes through zero.

Thus one can use analytic expressions for adiabatic curvature
perturbations based on slow-roll approximations in order to constrain
the model parameters of the potential.
Note that the number of e-folds, $\zk$, after the comoving wave-number $k$
leaves
the Hubble radius during inflation satisfies
\beq
  \frac{k}{k_f}=e^{-\lmk 1-\beta^2/2 \rmk\zk}
  (2\zk)^{n/2},~~~~~\mbox{for}~~ 1 \ll \zk \lesssim 60.
\eeq
We can then express the amplitude of curvature perturbation on scale
$l=2\pi/k$ as\cite{SY}
\beqa \Phi(l)&=&\lkk 1+\frac{2}{3(1+w)}\rkk^{-1}
\frac{\sqrt{2k^3\langle |C_1|^2 \rangle}}{2\pi} \nonumber \\
&=&\lkk 1+\frac{2}{3(1+w)}\rkk^{-1}
\frac{\kappa^2 e^{\beta^2\zk/2}}{2\pi}
\sqrt{\frac{\lambda_{2n}}{6n}
\lmk \frac{4n\zk}{\kappa^2}\rmk^{n}}
\lkk e^{-\beta^2\zk/4}
\sqrt{\frac{\zk}{n}}
+\frac{1}{\beta}(1-e^{-\beta^2\zk/2}) \rkk,
\eeqa
which is valid until the second horizon crossing after inflation and
also for $1 \ll \zk~\lsim~60$. Here $w$ is the ratio of the pressure to the
energy density.

Since the large-angular-scale anisotropy of background radiation due to
the Sachs-Wolfe effect is given by $\delta T/T = \Phi/3$,
we can normalize the value of $\lambda_{2n}$ by the COBE
normalization \cite{LR99}.  For $\beta=0.034$, since this gives the relation,
\beqa 
\frac{\delta T}{T} &=& \frac{1}{3}\Phi (\zk \simeq 60) \simeq \lnk
\begin{array}{lll} 9 \lambda_2/m_{\rm pl}
 \equiv & 9 m/m_{\rm pl} &
\mbox{~~~~~for $n=1$}\\
  28 \sqrt{\lambda_4}&~ &
  \mbox{~~~~~for $n=2$}\\
  \end{array} \right. \\
  &=& 1.1\times 10^{-5},
\eeqa
one finds
\beqa m&=& 1\times 10^{13}~\mbox{GeV},~~~~~n=1, \\
  \lambda_4 &=& 2\times 10^{-13},~~~~~~~~~~n=2,
\eeqa
which is not much different from the values obtained assuming the Einstein
gravity \cite{Salopek}.
Since the behavior of the system approaches to that in the
Einstein gravity as we increase $\omega$, we can conclude that in
Brans-Dicke theory, model
parameters of chaotic inflation should take the same value as
in the Einstein gravity.

Next we consider new inflation with a potential
\beq
  V(\phi)=V_0 -\frac{\lambda}{4}\phi^4,
\eeq
for which we find,
\beq
  \kappa^2\int_{\phi_f}^{\phi}\frac{V(\varphi)}{V'(\varphi)}d\varphi
\cong \frac{\kappa^2V_0}{2\lambda}\lmk \frac{1}{\phi^2}
- \frac{1}{\phi_f^2}\rmk \cong \frac{\kappa^2}{2\lambda\phi^2}
\simeq z.
\eeq
We can again express the amplitude of curvature fluctuation as a
function of $\zk$, which is now related with $k$
as,
\beq
\frac{k}{k_f}=e^{-\lmk 1-\beta^2/2 \rmk\zk},
\eeq
\beq
\Phi(l)=\lkk 1+\frac{2}{3(1+w)}\rkk^{-1}
\lkk e^{\beta^2\zk/4}
\sqrt{\frac{\lambda}{3}}(2\zk)^{3/2} +
\kappa H_f \frac{e^{\beta^2\zk/2}}{\beta}
(1-e^{-\beta^2\zk/2})\rkk,
\eeq
with $H_f\equiv \sqrt{\kappa^2V_0/3}$.
Taking $\beta=0.034$ again, it predicts the amplitude of
$\delta T/T$ to be compared with COBE data as
\beq
\frac{\delta T}{T}(\zk \simeq 60)
\simeq 30 \sqrt{\lambda} +0.21 \frac{H_f}{\mpl}.  \label{newinfdT}
\eeq
Since $H_f$ should  also satisfy
\beq
\frac{H_f}{m_{\rm pl}}~\lsim~10^{-5},
\eeq
to suppress long-wave gravitational
radiation of quantum origin \cite{GW}, we obtain
\beq
\lambda~\lsim~1\times 10^{-13},
\eeq
{}from Eq.~(\ref{newinfdT}).  Again its amplitude is
practically no different from the case of the Einstein gravity.

%
\subsection{Non-minimally coupled scalar field case}
%

Let us first briefly review the single-field inflationary scenario
with a non-minimally coupled scalar field ($\xi R\phi^2/2)$.
In chaotic inflationary models, Futamase and Maeda \cite{FM} found that the
non-minimal
coupling is constrained as $|\xi|~\lsim~10^{-3}$ in the quadratic
potential, by the requirement of sufficient amount of
inflation\footnote{This constraint is loosened by considering topological
inflation, see \cite{SY99}.}.
In the quartic potential, such a constraint is absent
for negative $\xi$, and as a bonus, the fine tuning problem
of the self-coupling $\lambda$ in the minimally coupled case
can be relaxed by considering large negative values of
$\xi$ \cite{Spo,SBB,FU}.
Several authors evaluated scalar and tensor perturbations generated during
inflation \cite{MS,Kaiser,Hwang99,KF} and preheating \cite{SB00} in this
model.  Since the system is reduced to the single-field case with a
modified inflaton potential by a conformal transformation, we cannot
expect nonadiabatic growth of ${\cal R}$ on large scales.

We shall proceed to the case of the non-minimally coupled $\psi$ field
with Eq.~(\ref{nonmin3}) in the presence of inflaton, $\phi$.
In this theory the evolution of field and metric perturbations was studied in
\cite{TY} in the Jordan frame.  It was found that ${\cal R}$ can grow
nonadiabatically during inflation on super-horizon scales for negative $\xi$.
Here we will show that similar results are obtained by the analysis in the
Einstein frame.

{}From Eqs.~(\ref{fieldper}), (\ref{SEPmet2}),
and (\ref{SEPzeta_m1}), we
obtain the following explicit solutions:
\begin{eqnarray}
\delta\chi=-4(C_1-C_3) \frac{\xi\psi}
{\sqrt{1-(1-6\xi)\xi\kappa^2\psi^2}},~~~~
\delta\phi=-\frac{V'(\phi)}{\kappa^2V(\phi)} \left[C_1
(1-\xi\kappa^2\psi^2)+C_3 (\xi\kappa^2\psi^2) \right],
\label{NONfield2}
\end{eqnarray}
\begin{eqnarray}
\Phi = -C_1\frac{\dot{H}}{H^2}-C_3\left[
\epsilon_{\psi}- (\xi\kappa^2\psi^2) \epsilon_{\phi} \right],
\label{NONmet2}
\end{eqnarray}
\begin{eqnarray}
{\cal R}= C_1-C_3 \frac{\epsilon_{\psi}- (\xi\kappa^2\psi^2)
\epsilon_{\phi}} {\epsilon_{\psi}+(1-\xi\kappa^2\psi^2
)\epsilon_{\phi}},
\label{NONzeta2}
\end{eqnarray}
with
\begin{eqnarray}
\epsilon_{\psi}=\frac{8(\xi\kappa\psi)^2} {1-(1-6\xi)\xi\kappa^2\psi^2},~~~~
\epsilon_{\phi}=\frac{1}{2\kappa^2} \left(\frac{V'(\phi)}{V(\phi)}\right)^2.
\label{NONepsilon}
\end{eqnarray}
When $\xi$ is negative, the coefficient of $\psi$ in the rhs
of Eq.~(\ref{varphiS}) is always positive,
which leads to the rapid growth of $\psi$
(and $\chi$).  Eq.~(\ref{NONfield2}) indicates that long wave $\delta\chi$
fluctuations are amplified with $|\psi|$ being increased.  This is due to the
fact that the effective mass of $\delta\chi$ becomes negative after horizon
crossing \cite{TY}, whose property is different from the JBD case.  In the
JBD case, $\delta\chi$ is almost constant during
inflation [see Eq.~(\ref{fieldper}) with $V_1=e^{-\beta\kappa\chi}$],
which restricts the nonadiabatic growth of large scale metric perturbations.
In contrast, in the present model, $\Phi$ and ${\cal R}$ exhibit strong
amplification due to the excitation of low momentum field perturbations
unless $|\psi|$ is initially very small.

The second terms in Eqs.~(\ref{NONmet2}) and (\ref{NONzeta2}) appear
in the presence of non-minimal coupling, whose contributions are
negligible when $|\xi|\kappa^2\psi^2 \ll 1$.
With the increase of $|\psi|$, however,
isocurvature perturbations are generated during inflation, which can lead
to nonadiabatic growth of $\Phi$ and ${\cal R}$.  When the second term in
Eq.~(\ref{NONmet2}) grows to of order the first term, the adiabatic mode
includes the isocurvature mode partially.  In this case one cannot
completely decompose adiabatic and isocurvature modes in the final results
by the expression, Eq.~(\ref{NONmet2}).

Let us consider the massive chaotic inflationary scenario
with initial conditions, $\phi_*=3m_{\rm pl}$
and $\chi_*=10^{-2}m_{\rm pl}$.
We plot in Fig.~2 the evolution of field perturbations for $\xi=-0.02$
in two cases, i.e., (i) solving directly the perturbed equations,
(\ref{perturb4})-(\ref{perturb3}), (ii) using the slow-roll solutions,
Eq.~(\ref{NONfield2}).  In spite of the rapid growth of field perturbations,
slow-roll analysis agrees reasonably well with full numerical results.
The enhancement of $\delta\chi_k$ fluctuations stimulates the amplification
of $\delta\phi_k$ fluctuations for $mt~\gsim~10$.  Inflationary period
ends around $mt \simeq 17$, after which the slow-roll results begin to
fail.  In Fig.~3, the evolution of $\Phi$ and ${\cal R}$ is depicted.  
We also plot the first and the second term in the rhs of 
Eq.~(\ref{NONmet2}), where
we denote $\Phi_1^{(s)}$ and $\Phi_2^{(s)}$, respectively. 
Although $\Phi$ is dominated by the $\Phi_1^{(s)}$ term in the initial stage,
$\Phi_2^{(s)}$ catches up $\Phi_1^{(s)}$ around $mt \simeq 7$ , 
after which the $\Phi_2^{(s)}$ term completely determines 
the evolution of $\Phi$.  We
find in Fig.~3 that slow-roll approximations are valid right up until the
end of inflation.  The growth of metric perturbations stops when the system
enters a reheating stage.

If we use the decomposition of Eq.~(\ref{SEPmet3}), the second term gives
negligible contribution to the gravitational potential around the end of
inflation.  In the early stage of inflation, however, its contribution is
comparable to the first term, in which case the isocurvature mode cannot
be completely separated from the adiabatic mode.  When fluctuations are
sufficiently amplified, it is inevitable that both adiabatic and
isocurvature modes mix each other with the growth of fluctuations, which
means that complete decomposition is difficult during the whole
inflationary stage.
\begin{figure}
\begin{center}
\singlefig{12cm}{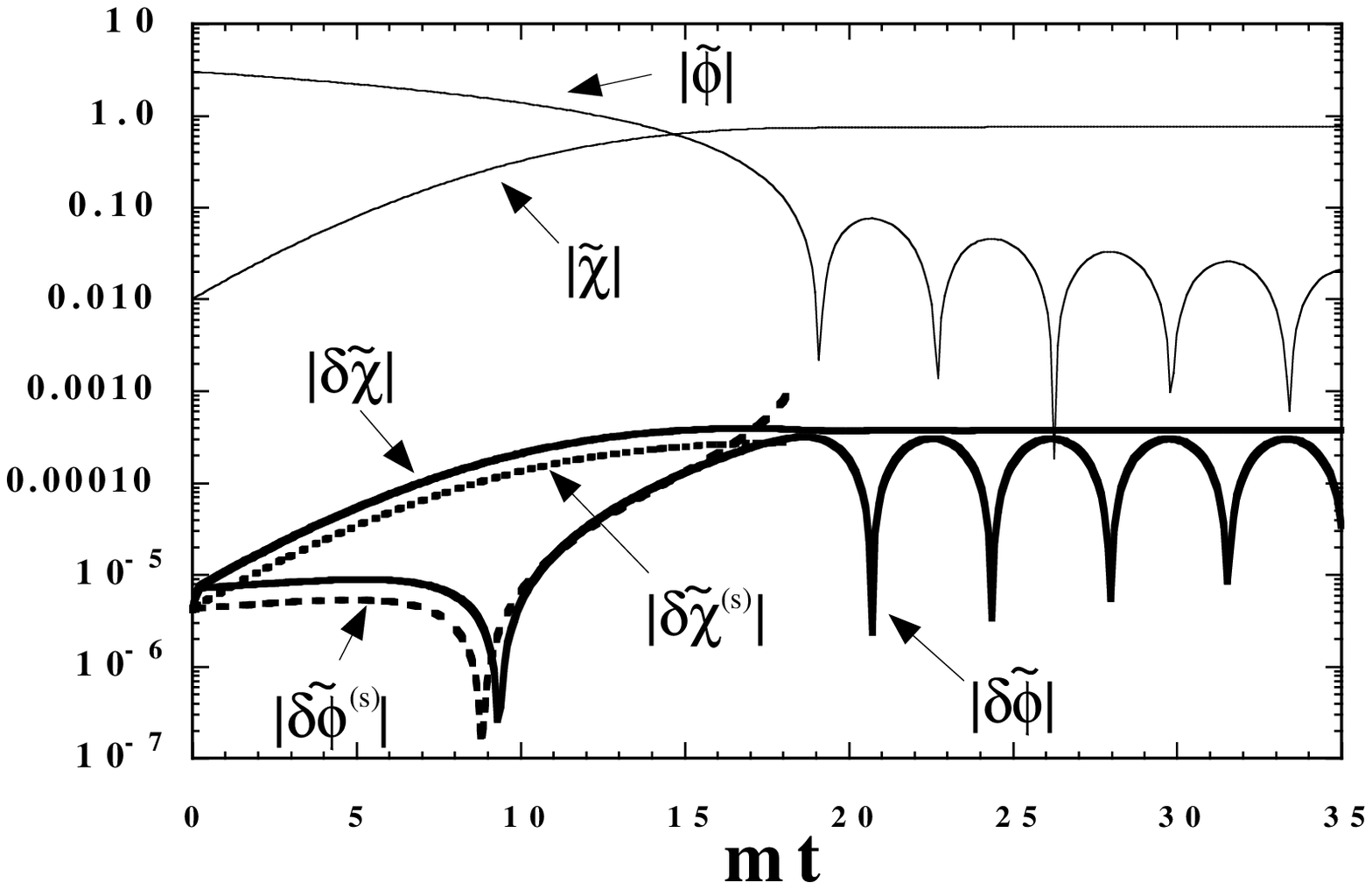}
\begin{figcaption}{Fig2}{12cm}
The evolution of background fields, $\tilde{\phi}=\phi/m_{\rm pl}$
and $\tilde{\chi}=\chi/m_{\rm pl}$, and large-scale field perturbations,
$\delta\tilde{\phi} \equiv k^{3/2} \delta\phi/m_{\rm pl}$
and $\delta\tilde{\chi} \equiv k^{3/2} \delta\chi/m_{\rm pl}$,
in the massive chaotic inflationary scenario
with a non-minimally coupled $\chi$ field for $\xi=-0.02$.
The initial values of scalar fields are chosen as
$\phi_*=3m_{\rm pl}$ and $\chi_*=10^{-2}m_{\rm pl}$.
The slow-roll results of Eq.~(\ref{NONfield2}) are also plotted,
where we denote $\delta\tilde{\phi}^{(s)}$ and $\delta\tilde{\chi}^{(s)}$ in the
 figure.
We find that slow-roll approximations are valid except in the final stage of
inflation ($mt>17$). The evolution of
$\delta\tilde{\phi}^{(s)}$ and $\delta\tilde{\chi}^{(s)}$ 
after inflation is not shown.
\end{figcaption}
\end{center}
\end{figure}

\begin{figure}
\begin{center}
\singlefig{12cm}{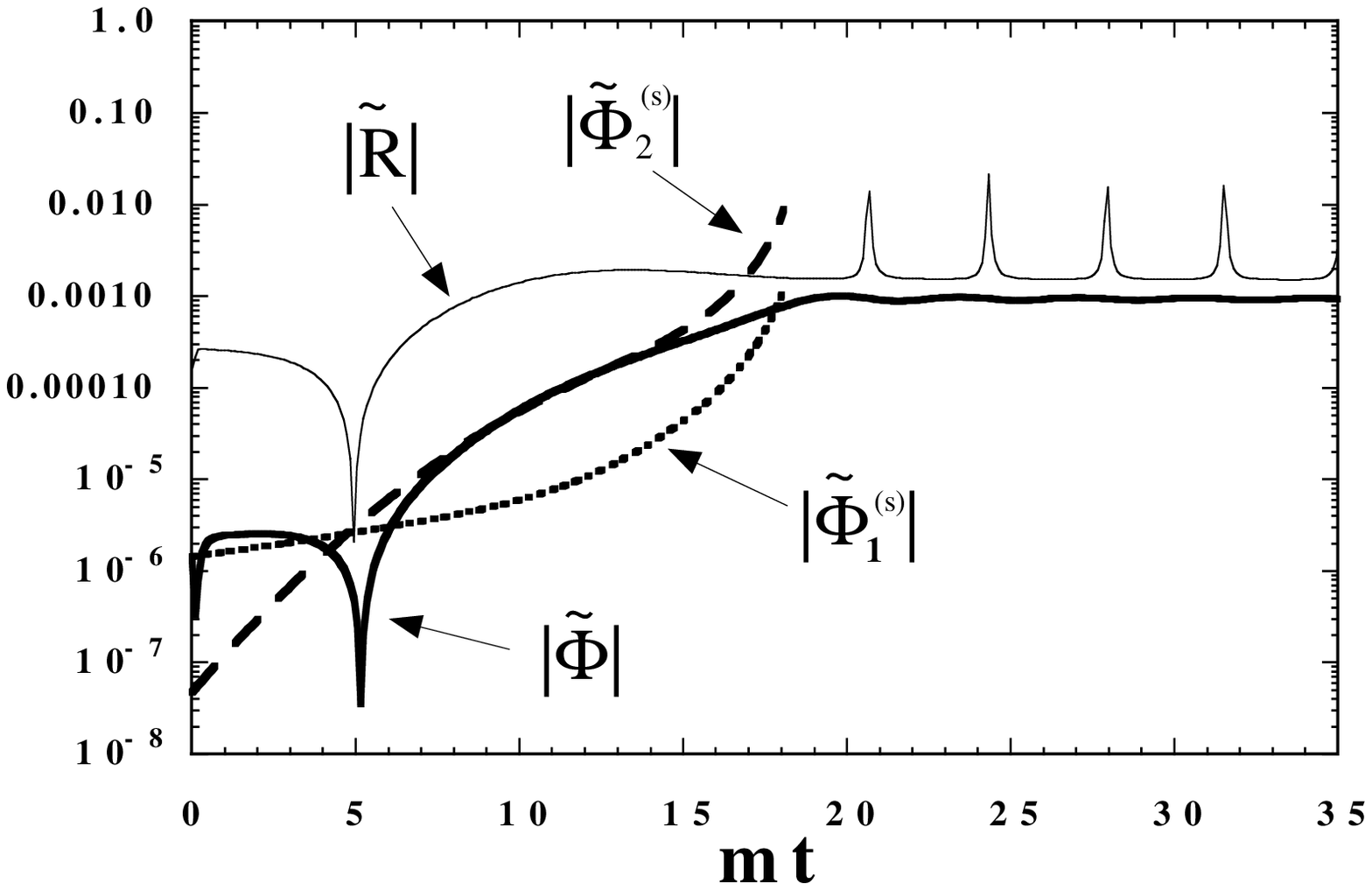}
\begin{figcaption}{Fig3}{12cm}
The evolution of the metric perturbation
$\tilde{\Phi} \equiv k^{3/2} \Phi$ and the curvature
perturbation $\tilde{\cal R}\equiv k^{3/2} {\cal R}$
with same parameters as in Fig.~2.
The amplification of field perturbations leads to the nonadiabatic 
growth of $\Phi$ and ${\cal R}$.
We also plot $\tilde{\Phi}_1^{(s)} \equiv k^{3/2} \Phi_1^{(s)}$
and $\tilde{\Phi}_2^{(s)} \equiv k^{3/2} \Phi_2^{(s)}$,
where $\Phi_1^{(s)}$ and $\Phi_2^{(s)}$ denote the first and the second
terms in Eq.~(\ref{NONmet2}), respectively.
$\Phi$ is mainly sourced by the
$\Phi_2^{(s)}$ term, after $\Phi_2^{(s)}$ 
catches up $\Phi_1^{(s)}$.  The evolution of 
$\Phi_1^{(s)}$ and $\Phi_2^{(s)}$ after inflation is 
not shown.
\end{figcaption}
\end{center}
\end{figure}

In Fig.~3 the curvature perturbation, ${\cal R}$, 
is nonadiabatically amplified sourced by the second term 
in Eq.~(\ref{NONzeta2}).  Whether this occurs or
not depends upon the strength of the coupling, $\xi$, and the initial
$\chi$.  When both are small and the second term in Eq.~(\ref{NONzeta2}) is
negligible relative to the $C_1$ term during inflation, we can regard the
first and second terms in Eq.~(\ref{NONzeta2}) as adiabatic and
isocurvature modes, respectively.  In the simulations of Figs.~2 and 3, we
take $\chi_*=10^{-2}m_{\rm pl}$, in which case numerical calculations
imply that the conservation of ${\cal R}$ is violated for $\xi~\lsim~-0.01$.
When $\xi~\lsim~-1$, strong amplification of ${\cal R}$ is inevitable 
even for very small values of $\chi$ far less than $m_{\rm pl}$.
For positive $\xi$, conservation of ${\cal R}$ is typically preserved due
to an exponential suppression of $\chi$ during inflation \cite{TY,BV,BPTV}.
Regarding detailed investigation about the observational constraints of the
strength of $\xi$, see \cite{TY} whose results are similar to
those in the analysis of the Einstein frame.

%
\subsection{Higher-dimensional theories}
%

In the higher-dimensional theory with Eq.~(\ref{FU2}), the kinetic term
takes a canonical form.  In this theory the condition, $D>1$, gives the
constraint, $\sqrt{2/3}<\beta<\sqrt{2}$, which is
different form the JBD theory with $\beta \ll 1$.  Larger values of $\beta$
correspond to the steep exponential potential of the $\chi$ field, which
leads to the rapid evolution toward the $\chi$ direction.
In this case inflaton decreases slowly relative to the $\chi$ field.
Then the expansion of the
universe is described by the power-law solution,
Eq.~(\ref{powerlaw}).

For example, in the polynomial inflaton potential, $V(\phi)
=\lambda_{2n}\phi^{2n}/(2n)$, classical trajectories of scalar fields
are given by Eq.~(\ref{C}) as
\begin{eqnarray}
C=\frac{\kappa}{\beta}\chi+\frac{\kappa^2}{4n}\phi^2.
\label{trajectry}
\end{eqnarray}
Differentiating Eq.~(\ref{trajectry}) with respect to $t$ yields
\begin{eqnarray}
\left| \frac{\dot{\chi}}{\dot{\phi}} \right|=
\frac{\beta\kappa}{2n}|\phi|=\frac{\sqrt{2\pi}\beta}{n}
\left| \frac{\phi}{m_{\rm pl}} \right|.
\label{dif}
\end{eqnarray}
This relation indicates that for the values of $\phi$ greater than $m_{\rm pl}$
with  $\beta$ and $n$ being of order unity, $|\dot{\chi}|$ is typically larger
than
 $|\dot{\phi}|$, in which case $\chi$ rapidly evolves along the exponential
potential.
The power-law inflation continues until $|\dot{\phi}|$ grows comparable
to $|\dot{\chi}|$, corresponding to $\phi/m_{\rm pl} \simeq n/(\sqrt{2\pi}\beta)
$.
After $\phi$ falls down this value, $\phi$ begins to evolve faster than
$\chi$ toward the local minimum at $\phi=0$ in the $\phi$ direction.
In this stage the system deviates from the power-law expansion, (\ref{powerlaw}).

{}From Eqs.~(\ref{SEPmet2}), (\ref{SEPmet3}),
(\ref{SEPzeta_m1}), and (\ref{SEPzeta}),
$\Phi$ and ${\cal R}$ evolve during inflation as
\begin{eqnarray}
\Phi =-C_1\frac{\dot{H}}{H^2}-C_3\frac{\beta^2}{2}
=-\tilde{C}_1\frac{\dot{H}}{H^2}-\tilde{C}_3
\frac{\beta^2-2\alpha_f\epsilon_{\phi}}{2(1+\alpha_f)},
\label{highermet}
\end{eqnarray}
\begin{eqnarray}
{\cal R} = C_1-C_3 \frac{\beta^2}{\beta^2+2\epsilon_{\phi}}
=\tilde{C}_1-\tilde{C}_3 \frac{\beta^2-2\alpha_f\epsilon_{\phi}}
{(1+\alpha_f)(\beta^2+2\epsilon_{\phi})},
\label{higherzeta}
\end{eqnarray}
where $\alpha_f=\beta^2/(2\epsilon_{\phi_f})$ is of order unity
for $\sqrt{2/3}<\beta<\sqrt{2}$.
Since the condition, $\beta^2 \gg \epsilon_{\phi}$,
holds during power law-inflation, the
curvature perturbation in this stage takes almost a constant value,
${\cal R} \simeq C_1-C_3=\tilde{C}_1-\tilde{C}_3/(1+\alpha_f)$.
As $|\dot{\phi}|$ grows relative to $|\dot{\chi}|$,
${\cal R}$ begins to evolve due to the change of $\epsilon_{\phi}$
in Eq.~(\ref{higherzeta}).
This corresponds to the stage where
deviations from power-law inflation become relevant.
When $\epsilon_{\phi}$ in Eq.~(\ref{higherzeta}) becomes
comparable to $\beta^2/2$ (i.e., $\alpha_f \simeq 1)$ around the
end of inflation, we have ${\cal R} \simeq C_1-C_3 \alpha_f/(1+\alpha_f)
=\tilde{C}_1$. After inflation ${\cal R}$ takes this conserved value.

%
\subsection{The $R^2$ theory with a non-minimally coupled
$\chi$ field}

In the $f(R)$ theories, effective potentials  do
not generally take separated forms, Eq.~(\ref{poten}), as found
in Eq.~(\ref{FU}).  Nevertheless we have closed form solutions,
Eqs.~(\ref{finalslow1})-(\ref{G2}), by which the evolution of
cosmological perturbations can be studied analytically.

Let us analyze the $R^2$ inflationary scenario with a
non-minimally coupled $\chi$ field as one example of
the $f(R)$ theory [see Eqs.~(\ref{R2FR}) and
(\ref{R2potential})].
Note that when $\chi=0$ the system has an effective potential,
\begin{eqnarray}
U=\frac{m_{\rm pl}^4}{(32\pi)^2\ab}
\left(1-e^{-(\sqrt{6}/3)\kappa\phi}\right)^2.
\label{R2pot}
\end{eqnarray}
The $\phi$ field defined by Eq.~(\ref{newphi}) plays the role of an inflaton
and leads to inflationary expansion of the Universe in the region,
$\phi~\gsim~m_{\rm pl}$ \cite{R2,TMT}. In the absence of the
non-minimally coupled $\chi$ field, the resulting spectrum of density
perturbations after the end of inflation was found in \cite{St83}
(using equations for perturbations of the FRW model for the Einstein gravity
with one-loop quantum corrections derived in \cite{St81}) and then
rederived in \cite{density_R2}. Here we study how the effect of
non-minimal coupling alters the adiabatic evolution of cosmological
perturbations in the single field case.

In the presence of
non-minimal coupling, Eq.~(\ref{G2}) is reduced to
\begin{eqnarray}
J=\frac{[1-(1-\xi\kappa^2\chi^2)e^{-(\sqrt{6}/3)\kappa\phi}]
(1-\xi\kappa^2\chi_*^2)}
{[1-(1-\xi\kappa^2\chi_*^2)e^{-(\sqrt{6}/3)\kappa\phi_*}]
(1-\xi\kappa^2\chi^2)}.
\label{G3}
\end{eqnarray}
Then Eqs.~(\ref{finalslow1}) and (\ref{finalslow2}) are integrated to give
\begin{eqnarray}
\delta\phi=-\frac{2\sqrt{6}(1-\xi\kappa^2\chi^2)
e^{-(\sqrt{6}/3)\kappa\phi}}{3\kappa
[1-(1-\xi\kappa^2\chi^2) e^{-(\sqrt{6}/3)\kappa\phi}]}
\left[C_1-C_3
\frac{\xi\kappa^2(\chi^2-\chi_*^2)}
{[1-(1-\xi\kappa^2\chi_*^2)e^{-(\sqrt{6}/3)\kappa\phi_*}]
(1-\xi\kappa^2\chi^2)}\right],
\label{R2field1}
\end{eqnarray}
\begin{eqnarray}
\delta\chi=-\frac{4\xi\chi}
{1-(1-\xi\kappa^2\chi^2) e^{-(\sqrt{6}/3)\kappa\phi}}
\left[C_1+C_3
\frac{1-(1-\xi\kappa^2\chi_*^2)e^{-(\sqrt{6}/3)\kappa\phi}}
{1-(1-\xi\kappa^2\chi_*^2)e^{-(\sqrt{6}/3)\kappa\phi_*}} \right],
\label{R2field2}
\end{eqnarray}
where $C_1=-\kappa^2Q_1$ and $C_3=-\kappa^2Q_3$.
Therefore $\Phi$ and ${\cal R}$ are expressed as
\begin{eqnarray}
\Phi = -C_1 \frac{\dot{H}}{H^2}-
\frac{C_3}{1-(1-\xi\kappa^2\chi_*^2) e^{-(\sqrt{6}/3)\kappa\phi_*}}
\left[ \epsilon_{\phi} \frac{\xi\kappa^2(\chi^2-\chi_*^2)}
{1-\xi\kappa^2\chi^2}-\epsilon_{\chi}
\left\{e^{(\sqrt{6}/3)\kappa\phi}-(1-\xi\kappa^2\chi_*^2)\right\}
\right],
\label{R2PHI}
\end{eqnarray}
\begin{eqnarray}
{\cal R}= C_1-\frac{C_3}{1-(1-\xi\kappa^2\chi_*^2)
e^{-(\sqrt{6}/3)\kappa\phi_*}}
\frac{\epsilon_{\phi}~\xi\kappa^2(\chi^2-\chi_*^2)-
\epsilon_{\chi}\left\{e^{(\sqrt{6}/3)\kappa\phi}-
(1-\xi\kappa^2\chi_*^2)\right\}}
{\epsilon_{\phi}+e^{(\sqrt{6}/3)\kappa\phi}\epsilon_{\chi}},
\label{R2zeta}
\end{eqnarray}
where $\epsilon_{\phi}$ and $\epsilon_{\chi}$ are defined by
\begin{eqnarray}
\epsilon_{\phi} &\equiv& \frac{1}{2\kappa^2}\left(
\frac{U_{,\phi}}{U} \right)^2=
\frac43 \left( \frac{(1-\xi\kappa^2\chi^2)
e^{-(\sqrt{6}/3)\kappa\phi}}
{1-(1-\xi\kappa^2\chi^2)e^{-(\sqrt{6}/3)\kappa\phi}}\right)^2,
\\
\epsilon_{\chi} &\equiv& \frac{1}{2\kappa^2}\left(
\frac{U_{,\chi}}{U} \right)^2= 8 \left( \frac{\xi\kappa\chi
e^{-(\sqrt{6}/3)\kappa\phi}}
{1-(1-\xi\kappa^2\chi^2)e^{-(\sqrt{6}/3)\kappa\phi}}\right)^2.
\label{R2slow}
\end{eqnarray}
In the absence of non-minimal coupling, one finds adiabatic results,
$\Phi=-C_1\dot{H}/{H^2}$ and ${\cal R}=C_1$.
In two-field inflation with a non-minimally coupled
$\chi$ field, the presence of isocurvature perturbations can lead to
nonadiabatic growth of $\Phi$ and ${\cal R}$ as found
in the second terms in Eqs.~(\ref{R2PHI}) and (\ref{R2zeta}).
Their contributions are negligible when the conditions,
$|\xi|\kappa^2\chi^2 \ll 1$ and $\epsilon_{\chi} \ll 1$,
holds during inflation.
The latter condition is similar to the former one
when $|\xi|~\lsim~1$.

{}From Eq.~(\ref{varphiS}), since  $\dot{\chi}$ is
approximately written as
\begin{eqnarray}
\dot{\chi}=-\frac{4\xi H}{1-(1-\xi\kappa^2\chi^2)
e^{-(\sqrt{6}/3)\kappa\phi}} \chi,
\label{R2dotchi}
\end{eqnarray}
the $\chi$ field exhibits exponential decrease for positive values of $\xi$.
For negative $\xi$, however, $\chi$ is exponentially amplified
during inflation, which means that the condition,
$|\xi|\kappa^2\chi^2 \ll 1$, can be violated.
The long-wave $\delta\chi$ fluctuation grows with the increase of $\chi$
as found in Eq.~(\ref{R2field2}).
On the other hand, the growth of $\delta\phi$ begins only when
the second term in Eq.~(\ref{R2field1})
becomes comparable to the first term.

\begin{figure}
\begin{center}
\singlefig{12cm}{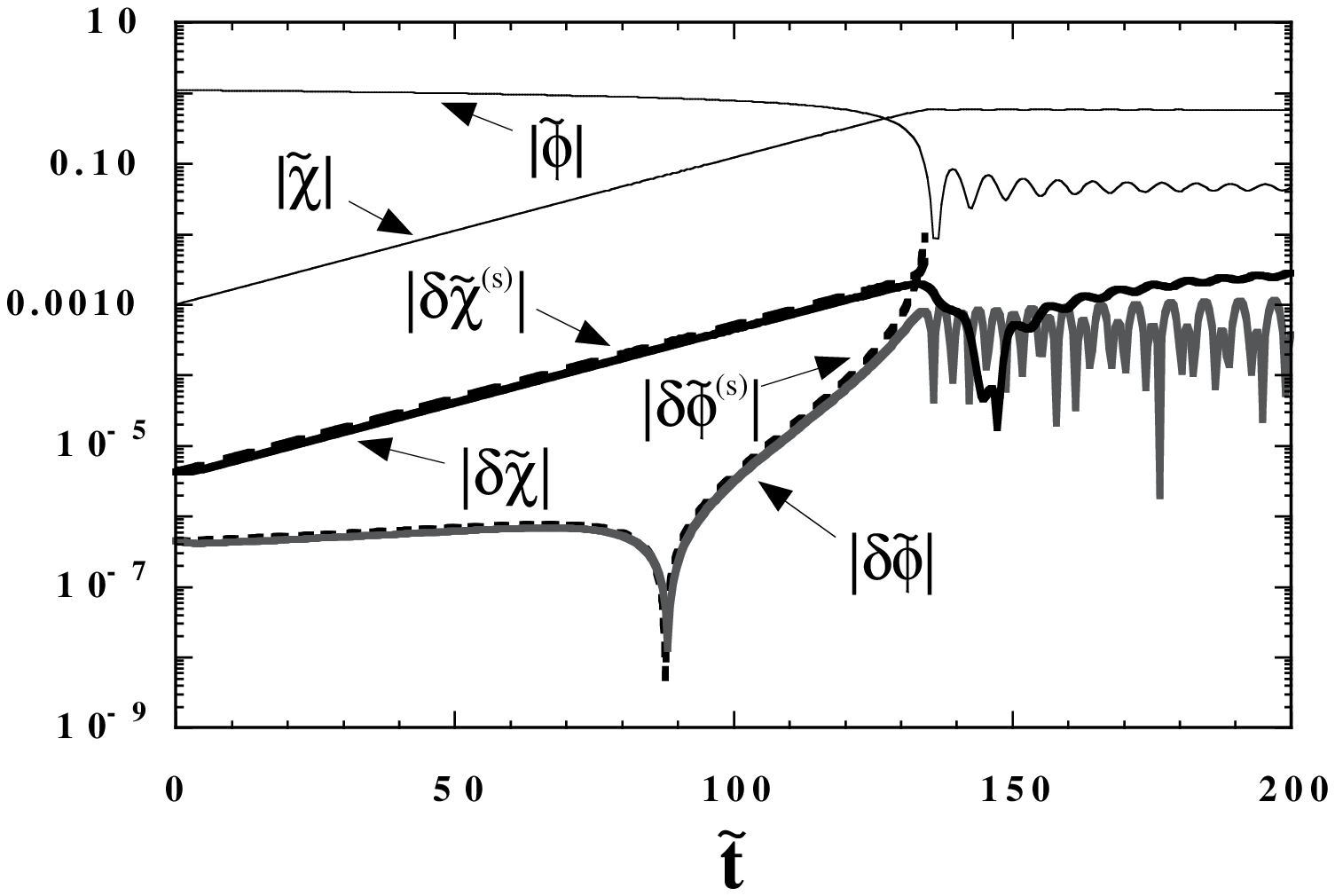}
\begin{figcaption}{Fig4}{12cm}
The evolution of background fields, $\tilde{\phi}=\phi/m_{\rm pl}$
and $\tilde{\chi}=\chi/m_{\rm pl}$, and long-wave field perturbations,
$\delta\tilde{\phi} \equiv k^{3/2} \delta\phi/m_{\rm pl}$ and
$\delta\tilde{\chi} \equiv k^{3/2} \delta\chi/m_{\rm pl}$ as
a function of time, $\bar{t} \equiv m_{\rm pl}t/\sqrt{96\pi \ab}$
in the $R^2$ inflationary scenario with a non-minimally coupled
$\chi$ field for $\xi=-0.025$.  The initial values of scalar fields are chosen
as
$\phi_*=1.1m_{\rm pl}$ and $\chi_*=10^{-3}m_{\rm pl}$.  We also plot the
slow-roll results, where we denote $\delta\tilde{\phi}^{(s)}$ and
$\delta\tilde{\chi}^{(s)}$ in the figure.  The evolution of
$\delta\tilde{\phi}^{(s)}$ and $\delta\tilde{\chi}^{(s)}$ 
after inflation is not shown.
\end{figcaption}
\end{center}
\end{figure}

\begin{figure}
\begin{center}
\singlefig{12cm}{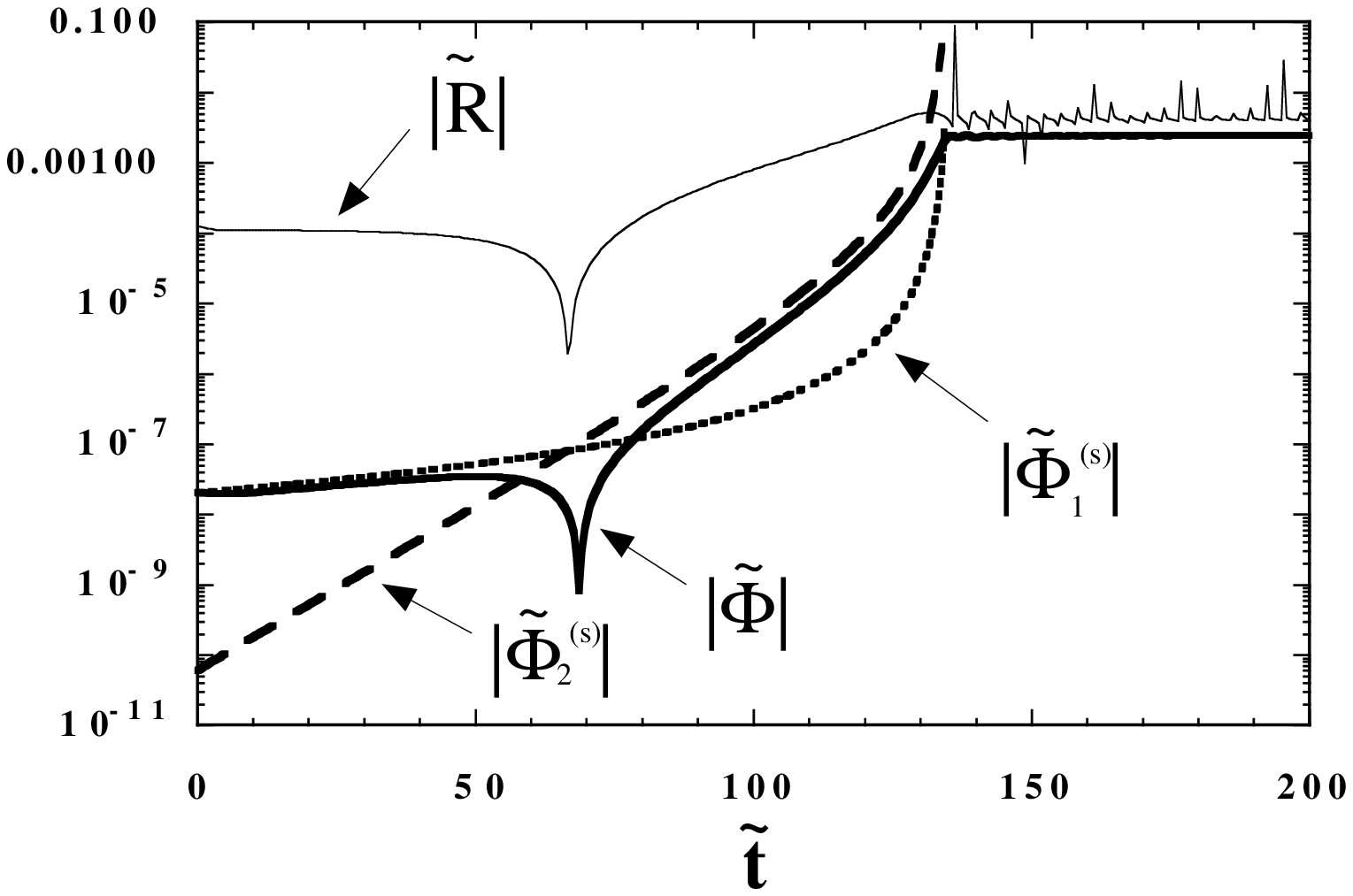}
\begin{figcaption}{Fig5}{12cm}
The evolution of the metric perturbation $\tilde{\Phi} \equiv k^{3/2}\Phi$
and the curvature perturbation $\tilde{\cal R} \equiv k^{3/2} {\cal R}$
in the $R^2$ inflationary scenario with
same parameters as in Fig.~4.
We also plot $\tilde{\Phi}_1^{(s)} \equiv k^{3/2} \Phi_1^{(s)}$
and $\tilde{\Phi}_2^{(s)} \equiv k^{3/2} \Phi_2^{(s)}$,
where $\Phi_1^{(s)}$ and $\Phi_2^{(s)}$ denote the first and the second
terms in Eq.~(\ref{R2PHI}), respectively.
The evolution of $\Phi_1^{(s)}$ and $\Phi_2^{(s)}$ 
after inflation is not shown.
\end{figcaption}
\end{center}
\end{figure}

We plot the evolution of $\delta\chi$, $\delta\phi$, and the slow-roll
results (\ref{R2field1})  and (\ref{R2field2})  for $\xi=-0.025$
with initial conditions,
$\phi_*=1.1m_{\rm pl}$ and $\chi_*=10^{-3}m_{\rm pl}$.
In this case inflation ends around $\tilde{t}\equiv
m_{\rm pl}t/\sqrt{96\pi \ab} \approx 130$ with e-foldings, $N \approx 63$.
Again the slow-roll analysis is quite reliable except around
the end of inflation.

In Fig.~5 we also depict the evolution of $\Phi$ and ${\cal R}$,
and the first ($=\Phi_1^{(s)}$) and second
($=\Phi_2^{(s)}$) terms in Eq.~(\ref{R2PHI}).  In the initial stage of
inflation where $\chi$ and $\delta\chi$ are not sufficiently amplified, the
gravitational potential is dominated by the $\Phi_1^{(s)}$ term, in which
case $\Phi_2^{(s)}$ may be regarded as the isocurvature mode.  However,
after $\tilde{t} \approx 70$ where $\Phi_2^{(s)}$ catches up
$\Phi_1^{(s)}$, we find in Fig.~5 that $\Phi_2^{(s)}$
mainly contributes to the
gravitational potential.  As is similar to the case of the chaotic
inflationary scenario with a non-minimally coupled $\chi$ field,
adiabatic and isocurvature modes mix each other
with the growth of the $\chi$ fluctuation.
If one defines the isocurvature mode as the one which gives
negligible contribution to the gravitational potential, it can not be
completely separated from the adiabatic mode during the whole stage of
inflation.

The conservation of ${\cal R}$ is typically violated when the  term
proportional to $C_3$ in Eq.~(\ref{R2zeta}) surpasses the one
proportional to $C_1$.  For the initial values,
$\phi_*=1.1m_{\rm pl}$ and $\chi_*=10^{-3}m_{\rm pl}$,
${\cal R}$ exhibits nonadiabatic growth for $\xi~\lsim~-0.02$.
Negative large non-minimal coupling such as $\xi~\lsim~-1$ leads to strong
amplification of ${\cal R}$ unless $\chi$ is initially very small.
Although we do not make detailed analysis here,
the basic property is quite similar to the case of the subsection B.
These results are also expected to hold for other inflationary models with a
non-minimally coupled $\chi$ field, since the scalar curvature is
proportional to the potential energy of inflaton which slowly decreases
during inflation.

 \section{Conclusions}

We have studied generation and evolution of adiabatic and
isocurvature perturbations during multi-field inflation in
generalized Einstein theories. Most of the generalized Einstein
theories recast to the Lagrangian (\ref{lagrangian}) in the
Einstein frame by conformal transformations. While behavior of
adiabatic perturbations is universal and is given by Eqs. (\ref{adia},
\ref{phiadia}) in the long-wave limit ($k\to 0$), isocurvature
perturbations behave differently depending on specific gravity
theories. Making use of slow-roll approximations, we have obtained
closed form solutions for all non-decaying field and metric perturbations
in the long-wave limit. The existence of isocurvature perturbations
may lead to significant variations of the curvature perturbations
${\cal R}$ and $\zeta$.

In this work we considered the following four gravity theories.

(1) The Jordan-Brans-Dicke theory with a Brans-Dicke field $\chi$ and
inflaton $\phi$.  Using the Brans-Dicke parameter constrained by
observations, the isocurvature mode in the gravitational potential $\Phi$ is
negligible relative to the adiabatic mode.  Therefore the variation of
${\cal R}$ is typically small in this theory.
In particular, for the quartic potential, ${\cal R}$ is conserved in the
slow-roll analysis.

(2) A non-minimally coupled scalar field $\chi$ in the presence of inflaton
$\phi$.  When the coupling $\xi$ is negative, $\chi$ and its long-wave
fluctuations exhibit exponential increase during inflation,
leading to the nonadiabatic amplification of $\Phi$ and ${\cal R}$
due to the existence of isocurvature perturbations.
Even in this case we find
that slow-roll analysis agrees well with full numerical results.
When field and metric perturbations are sufficiently amplified,
adiabatic and isocurvature modes of the gravitational potential mix
each other.

(3) Higher-dimensional Kaluza-Klein theory with dilaton $\chi$ and
inflaton $\phi$.  In this theory the inflationary period can be divided
into  two stages: the first is the power-law inflationary stage
where $\chi$ evolves along the exponential potential and
the second is the deviation from power-law inflation
due to the rapid evolution of $\phi$ around the end
of inflation.
In the former stage ${\cal R}$ is nearly constant but its change
occurs at the transition between two stages.

(4) $R^2$ theory with a non-minimally coupled scalar field $\chi$.
This system has an additional scalar field $\phi$ playing the role of
inflaton after conformal transformations.  Although this theory has
a coupled effective potential which is different from the above theories
(1)-(3), we have integrated forms of long-wave field and metric
perturbations by slow-roll analysis, which is found to be quite reliable
right up until the end of inflation.
Negative non-minimal coupling again leads
to the nonadiabatic growth of $\Phi$ and ${\cal R}$, in which case
complete decomposition between adiabatic and isocurvature modes
is difficult.

While we analyzed generalized Einstein theories involving
two scalar fields,
there exist other multi-field inflationary scenarios such as hybrid
inflation \cite{hybrid} and models of two interacting
scalar fields \cite{KL87}. During preheating after inflation,
there has been growing interest
about the evolution of cosmological perturbations for the simple
two-field model with potential
$U(\phi,\chi)=\lambda \phi^4/4+
g^2\phi^2\chi^2/2$ \cite{BV,selfpre}.
It is certainly of interest to constrain realistic multi-field
inflationary models based on particle physics using density perturbations
generated during inflation, together with constraints by gravitinos
\cite{gravitinos} and possible primordial black hole
over-production during preheating \cite{PBH}.

\section*{ACKNOWLEDGEMENTS}
AS is thankful to Profs. Katsuhiko Sato and Masahiro Kawasaki for
hospitality in RESCEU, the University of Tokyo. He was also partially
supported by the Russian Fund for Fundamental Research, grants 99-02-16224
and 00-15-96699.
ST thanks Bruce A. Bassett, Kei-ichi Maeda, Naoshi Sugiyama,
Atsushi Taruya, Takashi Torii, Hiroki Yajima, 
and David Wands for useful discussions.
He is also thankful for financial support from the JSPS (No. 04942).
The work of JY was partially supported by the Monbukagakusho
Grant-in-Aid, Priority Area ``Supersymmetry and Unified Theory of
Elementary Particles''(\#707) and the Monbukagakusho Grant-in-Aid for
Scientific Research No.\ 112440063.


\end{document}